# A dichotomy theorem for conservative general-valued CSPs

Vladimir Kolmogorov

University College London

v.kolmogorov@cs.ucl.ac.uk

**Abstract**

We study the complexity of valued constraint satisfaction problems (VCSP). A problem from VCSP is characterised by a *constraint language*, a fixed set of cost functions over a finite domain. An instance of the problem is specified by a sum of cost functions from the language and the goal is to minimise the sum. We consider the case of so-called *conservative* languages; that is, languages containing all unary cost functions, thus allowing arbitrary restrictions on the domains of the variables. We prove a Schaefer-like dichotomy theorem for this case: if all cost functions in the language satisfy a certain condition (specified by a complementary combination of *STP and MJN multimorphisms*) then any instance can be solved in polynomial time by the algorithm of Kolmogorov and Živný (arXiv:1008.3104v1), otherwise the language is NP-hard. This generalises recent results of Takhanov (STACS'10) who considered $\{0, \infty\}$-valued languages containing additionally all finite-valued unary cost functions, and Kolmogorov and Živný (arXiv:1008.1555v1) who considered *finite-valued* conservative languages.

## 1 Introduction

The constraint satisfaction problem is a central generic problem in computer science. It provides a common framework for many theoretical problems as well as for many real-life applications, see [23] for a nice survey. An instance of the *constraint satisfaction problem* (CSP) consists of a collection of variables which must be assigned values subject to specified constraints. CSP is known to be equivalent the problem of evaluating conjunctive queries on databases [27], and to the homomorphism problem for relations structures [20].

An important line of research on the CSP is to identify all tractable cases; that is, cases that are recognisable and solvable in polynomial time. Most of this work has been focused on one of the two general approaches: either identifying structural properties of the way constraints interact which ensure tractability no matter what forms of constraint are imposed [18], or else identifying forms of constraint which are sufficiently restrictive to ensure tractability no matter how they are combined [9, 20].

The first approach has been used to characterise all tractable cases of bounded-arity CSPs: the *only* class of structures which ensures tractability (subject to a certain complexity theory assumption, namely FPT $\neq$ W[1]) are structures of bounded tree-width modulo homomorphic equivalence [16, 22]. The second approach has led to identifying certain algebraic properties known as polymorphisms [24] which are necessary for a set of constraint types to ensure tractability. A set of constraint types which ensures tractability is called a *tractable constraint language*.

Schaefer in his seminal work [34] gave a complete complexity classification of Boolean constraint languages. The algebraic approach based on polymorphisms [25] has been so far the most successful tool in generalising Schaefer's result to languages over a 3-element domain [10], conservative languages [8], languages comprising of a single binary relation without sources and sinks [5] (see also [3]), and languages comprising of a single binary relation that is a special triad [4]. The algebraic approach has also been essential in characterising the power of local consistency [2] and the "few subpowers property" [6], the two main tools known for solving tractable CSPs. The ultimate goal in this line of research is to answer the *Dichotomy Conjecture* of Feder and Vardi, which states that every constraint language is either tractable or NP-hard [20]. We remark that there are other approaches to the dichotomy conjecture; see, for instance, [23] for a nice survey of Hell and Nešetřil, and [30] for a connection between the Dichotomy Conjecture and probabilistically checkable proofs.

Since in practice many constraint satisfaction problems are over-constrained, and hence have no solution, or are under-constrained, and hence have many solutions, *soft* constraint satisfaction problems have been studied [17]. In an instance of the soft CSP, every constraint is associated with a function (rather than a relation as in the CSP) which represents preferences among different partial assignments, and the goal is to find the best assignment. Several very general soft CSP frameworks have been proposed in the literature [35, 7]. In this paper we focus on one of the very general frameworks, the *valued* constraint satisfaction problem (VCSP) [35].

Similarly to the CSP, an important line of research on the VCSP is to identify tractable cases which are recognisable in polynomial time. Is is well known that structural reasons for tractability generalise to the VCSP [17]. In the case of language restrictions, only a few conditions are known to guarantee tractability of a given set of valued constraints [13, 12].

**Related work** The problem of characterising complexity of different classes of languages has received significant attention in the literature. For some classes researchers have established a Schaefer-like dichotomy theorem of the following form: if language $\Gamma$ admits certain *polymorphisms* or *multimorphisms* then it is tractable, otherwise it is NP-hard. Some of these classes are listed below:

- Class of Boolean languages, i.e. the case a 2-element domain (Cohen *et al.* [13]).

- Class of crisp languages with a 3-element domain (Bulatov [10]).

- Class of crisp conservative languages (Bulatov [8]).

- Class of conservative $\{0, 1\}$-valued cost function languages, with some generalisations (Deineko *et al.* [19])

- Class of crisp languages $\Gamma$ including additionally all finite-valued unary cost functions (Takhanov [36]).

- Class of crisp languages $\Gamma$ including additionally a certain subset of finite-valued unary cost functions (Takhanov [37]).

- Class of finite-valued conservative languages (Kolmogorov and Živný [28]).

Other related work includes the work of Creignou *et al.* who studied various generalisations of the CSP to optimisation problems over Boolean domains [14], see also [15, 26]. Raghavendra [32] and Raghavendra and Steurer [33] have shown how to optimally approximate any finite-valued VCSP.

**Contributions** We prove a dichotomy theorem for general-valued conservative languages: if a conservative language $\Gamma$ admits a complementary combination of *STP and MJN multimorphisms* then it is tractable, otherwise $\Gamma$ is NP-hard. Thus, we add another class to the list above generalising classes considered in [36] and [28].

We prove this result by constructing a (partial) *MJN multimorphism*, assuming that $\Gamma$ satisfies certain properties. The dichotomy theorem is then obtained with the help of two previous papers by Kolmogorov and Živný [28, 29]. The first one [28] proved the existence of a (partial) *STP multimorphism*, assuming that $\Gamma$ satisfies certain properties. The second paper [29] proved that a combination of an STP and MJN multimorphisms ensures tractability.

Our proof exploits recent results of Takhanov [36] and Kolmogorov and Živný [28], who showed the existence of respectively a majority polymorphism and an STP multimorphism under certain conditions.

## 2 Background and notation

We denote by $\mathbb{R}_+$ the set of all non-negative real numbers. We denote $\overline{\mathbb{R}}_+ = \mathbb{R}_+ \cup \{\infty\}$ with the standard addition operation extended so that for all $a \in \mathbb{R}_+$, $a + \infty = \infty$. Members of $\overline{\mathbb{R}}_+$ are called *costs*.

Throughout the paper, we denote by $D$ any fixed finite set, called a *domain*. Elements of $D$ are called *domain values* or *labels*.

A function $f$ from $D^m$ to $\overline{\mathbb{R}}_+$ will be called a *cost function* on $D$ of *arity* $m$. If the range of $f$ lies entirely within $\mathbb{R}$, then $f$ is called a *finite-valued* cost function. If the range of $f$ is $\{0, \infty\}$, then $f$ is called a *crisp* cost function. If the range of a cost function $f$ includes non-zero finite costs and infinity, we emphasise this fact by calling $f$ a *general-valued* cost function. Let $f : D^m \to \overline{\mathbb{R}}_+$ be an $m$-ary cost function $f$. We denote $\text{dom} f = \{ \boldsymbol{x} \in D^m \mid f(\boldsymbol{x}) < \infty \}$ to be the effective domain of $f$. The argument of $f$ is called an *assignment* or a *labeling*. Functions $f$ of arity $m = 2$ are called *binary*.

A *language* is a set of cost functions with the same set $D$. Language $\Gamma$ is called finite-valued (crisp, general-valued respectively) if all cost functions in $\Gamma$ are finite-valued (crisp, general-valued respectively). A language $\Gamma$ is *Boolean* if $|D| = 2$.

**Definition 1.** *An instance $\mathcal{I}$ of the* valued constraint satisfaction problem *(VCSP) is a function $D^V \to \overline{\mathbb{R}}_+$ given by*

$$\text{Cost}_\mathcal{I}(\boldsymbol{x}) = \sum_{t \in T} f_t \left( x_{i(t,1)}, \ldots, x_{i(t,m_t)} \right)$$

*It is specified by a finite set of nodes $V$, finite set of terms $T$, cost functions $f_t : D^{m_t} \to \overline{\mathbb{R}}_+$ or arity $m_t$ and indices $i(t, k) \in V$ for $t \in T$, $k = 1, \ldots, m_t$. A* solution *to $\mathcal{I}$ is an assignment $\boldsymbol{x} \in D^V$ with the minimum cost.*

We denote by $\text{VCSP}(\Gamma)$ the class of all VCSP instances whose terms $f_t$ belong to $\Gamma$. A finite language $\Gamma$ is called *tractable* if $\text{VCSP}(\Gamma)$ can be solved in polynomial time, and *intractable* if $\text{VCSP}(\Gamma)$ is NP-hard. An infinite language $\Gamma$ is tractable if every finite subset $\Gamma' \subseteq \Gamma$ is tractable, and intractable if there is a finite subset $\Gamma' \subseteq \Gamma$ that is intractable.

Intuitively, a language is conservative if one can restrict the domain of any variable to an arbitrary subset of the domain.

**Definition 2.** *Language $\Gamma$ is called* conservative *if $\Gamma$ contains all finite-valued unary cost functions $u : D \to \mathbb{R}_+$.*

We remark that in the crisp case, conservative languages are defined differently in the literature: a crisp language is called conservative if it contains all possible $\{0, \infty\}$-valued unary cost functions [8]. In this paper we always use Definition 2, unless explicitly referring to the crisp case.

**Definition 3.** *A mapping $F : D^k \to D$, $k \geq 1$ is called a* polymorphism *of a cost function $f : D^m \to \overline{\mathbb{R}}_+$ if*

$$F(\boldsymbol{x}_1, \ldots, \boldsymbol{x}_k) \in \text{dom} f \qquad \forall \boldsymbol{x}_1, \ldots, \boldsymbol{x}_k \in \text{dom} f$$

*where $F$ is applied component-wise. $F$ is a polymorphism of a language $\Gamma$ if $F$ is a polymorphism of every cost function in $\Gamma$.*

Multimorphisms [13] are generalisations of polymorphisms. To make the paper easier to read, we only define binary and ternary multimorphisms as we will not need multimorphisms of higher arities.

**Definition 4.** *Let $\langle \sqcap, \sqcup \rangle$ be a pair of operations, where $\sqcap, \sqcup : D \times D \to D$, and let $\langle F_1, F_2, F_3 \rangle$ be a triple of operations, where $F_i : D \times D \times D \to D$, $1 \leq i \leq 3$.*

- *Pair $\langle \sqcap, \sqcup \rangle$ is called a (binary)* multimorphism *of cost function $f : D^m \to \overline{\mathbb{R}}_+$ if*

$$f(\boldsymbol{x} \sqcap \boldsymbol{y}) + f(\boldsymbol{x} \sqcup \boldsymbol{y}) \leq f(\boldsymbol{x}) + f(\boldsymbol{y}) \qquad \forall \boldsymbol{x}, \boldsymbol{y} \in \mathrm{dom} f \tag{1}$$

   *where operations $\sqcap, \sqcup$ are applied component-wise. $\langle \sqcap, \sqcup \rangle$ is a multimorphism of language $\Gamma$ if $\langle \sqcap, \sqcup \rangle$ is a multimorphism of every $f$ from $\Gamma$.*

- *Triple $\langle F_1, F_2, F_3 \rangle$ is called a (ternary)* multimorphism *of cost function $f : D^m \to \overline{\mathbb{R}}_+$ if*

$$f(F_1(\boldsymbol{x}, \boldsymbol{y}, \boldsymbol{z})) + f(F_2(\boldsymbol{x}, \boldsymbol{y}, \boldsymbol{z})) + f(F_3(\boldsymbol{x}, \boldsymbol{y}, \boldsymbol{z})) \leq f(\boldsymbol{x}) + f(\boldsymbol{y}) + f(\boldsymbol{z}) \quad \forall \boldsymbol{x}, \boldsymbol{y}, \boldsymbol{z} \in \mathrm{dom} f \tag{2}$$

   *where operations $F_1, F_2, F_3$ are applied component-wise. $\langle F_1, F_2, F_3 \rangle$ is a multimorphism of language $\Gamma$ if $\langle F_1, F_2, F_3 \rangle$ is a multimorphism of every $f$ from $\Gamma$.*

- *Pair $\langle \sqcap, \sqcup \rangle$ is called* conservative *if $\{a \sqcap b, a \sqcup b\} = \{a, b\}$ for all $a, b \in D$. Operation $F_i$ is called conservative $F_i(a, b, c) \in \{a, b, c\}$ for all $a, b, c \in D$.*

- *Pair $\langle \sqcap, \sqcup \rangle$ is called a* symmetric tournament pair (STP) *if it is conservative and both operations $\sqcap, \sqcup$ are commutative, i.e. $a \sqcap b = b \sqcap a$ and $a \sqcup b = b \sqcup a$ for all $a, b \in D$.*

- *An operation $\mathtt{Mj} : D^3 \to D$ is called a* majority operation *if for every tuple $(a, b, c) \in D^3$ with $|\{a, b, c\}| = 2$ operation $\mathtt{Mj}$ returns the unique majority element among $a, b, c$ (that occurs twice). An operation $\mathtt{Mn} : D^3 \to D$ is called a* minority operation *if for every tuple $(a, b, c) \in D^3$ with $|\{a, b, c\}| = 2$ operation $\mathtt{Mn}$ returns the unique minority element among $a, b, c$ (that occurs once).*

- *Triple $\langle \mathtt{Mj}_1, \mathtt{Mj}_2, \mathtt{Mn}_3 \rangle$ is called an* MJN *if $\mathtt{Mj}_1, \mathtt{Mj}_2$ are (possibly different) majority operations, $\mathtt{Mn}_3$ is a minority operation, and each each operation $\mathtt{Mj}_1, \mathtt{Mj}_2, \mathtt{Mj}_3 : D^3 \to D$ is conservative.*

We say that $\langle \sqcap, \sqcup \rangle$ is a multimorphism of language $\Gamma$, or $\Gamma$ admits $\langle \sqcap, \sqcup \rangle$, if all cost functions $f \in \Gamma$ satisfy (1). Using a polynomial-time algorithm for minimising submodular functions, Cohen *et al.* have obtained the following result:

**Theorem 5** ([12]). *If a language $\Gamma$ admits an STP, then $\Gamma$ is tractable.*

The existence of an MJN multimorphism also leads to tractability. This was shown for a specific choice of an MJN by Cohen *et al.* [13], and for arbitrary MJNs by Kolmogorov and Živný [29]:

**Theorem 6** ([29]). *If a language $\Gamma$ admits an MJN, then $\Gamma$ is tractable.*

In fact, Kolmogorov and Živný considered a more general class of functions that includes two classes above as special cases; the exact formulation is given in the next section.

**Expressibility** Finally, we define the important notion of expressibility, which captures the idea of introducing auxiliary variables in a VCSP instance and the possibility of minimising over these auxiliary variables. (For crisp languages, this is equivalent to *implementation* [15].)

**Definition 7.** *A cost function* $f : D^m \to \overline{\mathbb{R}}_+$ *is* expressible *over a language* $\Gamma$ *if there exists an instance* $\mathcal{I} \in \mathsf{VCSP}(\Gamma)$ *with the set of nodes* $V = \{1, \ldots, m, m+1, \ldots, m+k\}$ *where* $k \geq 0$ *such that*

$$f(\boldsymbol{x}) = \min_{\boldsymbol{y} \in D^k} Cost_{\mathcal{I}}(\boldsymbol{x}, \boldsymbol{y}) \qquad \forall \boldsymbol{x} \in D^m$$

*We define* $\Gamma^*$ *to be the* expressive power *of* $\Gamma$*; that is, the set of all cost functions* $f$ *such that* $f$ *is expressible over* $\Gamma$.

The importance of expressibility is in the following result:

**Theorem 8** ([13]). *For any language* $\Gamma$*,* $\Gamma$ *is tractable iff* $\Gamma^*$ *is tractable.*

It is easy to observe and well known that any polymorphism (multimorphism) of $\Gamma$ is also a polymorphism (multimorphism) of $\Gamma^*$ [13].

## 3 Our results

To formulate our result, we will first recall the following definition from Kolmogorov and Živný [28]. Given a conservative language $\Gamma$, let $G_\Gamma = (P, E)$ be the graph with the set of nodes $P = \{(a,b) | a, b \in D, a \neq b\}$ and the set of edges $E$ defined as follows: there is an edge between $(a, b) \in P$ and $(a', b') \in P$ iff there exists binary cost function $f \in \Gamma^*$ such that

$$f(a, a') + f(b, b') > f(a, b') + f(b, a'), \quad (a, b'), (b, a') \in \mathrm{dom}\, f \tag{3}$$

Note that $G_\Gamma$ may have self-loops. For node $(a, b) \in P$ we denote the self-loop by $\{(a,b), (a,b)\}$. We say that edge $\{(a, b), (a', b')\} \in E$ is *soft* if there exists binary $f \in \Gamma^*$ satisfying (3) such that at least one of the assignments $(a, a'), (b, b')$ is in $\mathrm{dom}\, f$. Edges in $E$ that are not soft are called *hard*.

We denote $M \subseteq P$ to be the set of vertices $(a, b) \in P$ without self-loops, and $\overline{M} = P - M$ to be the complement of $M$.

**Lemma 9** ([28]). *Graph* $G_\Gamma$ *satisfies the following properties:*

(a) *For each* $(a, b) \in P$*, nodes* $(a, b)$ *and* $(b, a)$ *are either both in* $M$ *or both in* $\overline{M}$.

(b) *There are no edges from* $M$ *to* $\overline{M}$.

(c) *Nodes* $(a, b) \in \overline{M}$ *do not have incident soft edges.*

We will write $\{a, b\} \in M$ to indicate that $(a, b) \in M$; this is consistent due to lemma 9(a). Similarly, we will write $\{a, b\} \in \overline{M}$ if $(a, b) \in \overline{M}$, and $\{a, b\} \in P$ if $(a, b) \in P$, i.e. $a, b \in D$ and $a \neq b$.

**Definition 10.** *Let* $\langle \sqcap, \sqcup \rangle$ *and* $\langle \mathtt{Mj}_1, \mathtt{Mj}_2, \mathtt{Mn}_3 \rangle$ *be binary and ternary operations respectively.*

- *Pair* $\langle \sqcap, \sqcup \rangle$ *is an STP on* $M$ *if* $\langle \sqcap, \sqcup \rangle$ *is conservative on* $P$ *and commutative on* $M$.

- *Triple* $\langle \mathtt{Mj}_1, \mathtt{Mj}_2, \mathtt{Mn}_3 \rangle$ *is an MJN on* $\overline{M}$ *if operations* $\mathtt{Mj}_1, \mathtt{Mj}_2, \mathtt{Mn}_3$ *are conservative and for each triple* $(a, b, c) \in D^3$ *with* $\{a, b, c\} = \{x, y\} \in \overline{M}$ *operations* $\mathtt{Mj}_1(a, b, c)$*,* $\mathtt{Mj}_2(a, b, c)$ *return the unique majority element among* $a, b, c$ *(that occurs twice) and* $\mathtt{Mn}_3(a, b, c)$ *returns the remaining minority element.*

The importance of this definition follows from the following two theorems of Kolmogorov and Živný [28, 29].

**Theorem 11** ([28]). *Let $\Gamma$ be a conservative language.*

(a) *If $G_\Gamma$ has a soft self-loop then $\Gamma$ is NP-hard.*

(b) *If $G_\Gamma$ does not have soft self-loops then $\Gamma$ does admits a pair $\langle \sqcup, \sqcap \rangle$ which is an STP on $M$ and satisfies additionally $a \sqcap b = a, a \sqcup b = b$ for $\{a, b\} \in \overline{M}$.*

Note, this theorem implies the dichotomy result for **finite-valued** languages: we then have $M = P$, so in the case (b) $\Gamma$ is tractable by theorem 5.

**Theorem 12** ([29]). *Suppose language $\Gamma$ admits an STP on $M$ and an MJN on $\overline{M}$, for some choice of $M \subseteq P$. Then $\Gamma$ is tractable.*

The main result of this paper is the following

**Theorem 13.** *Let $\Gamma$ be a conservative language. If $G_\Gamma$ does not have soft self-loops and admits a majority polymorphism then $\Gamma$ admits an MJN on $\overline{M}$. Otherwise $\Gamma$ is NP-hard.*

When combined with theorems 11 and 12, it gives the dichotomy result for conservative languages:

**Corollary 14.** *If a conservative language $\Gamma$ admits an STP on $M$ and an MJN on $\overline{M}$ for some $M \subseteq P$ then $\Gamma$ is tractable. Otherwise $\Gamma$ is NP-hard.*

## 4 Proof of theorem 13

Let $\bar{\Gamma}$ be the language obtained from $\Gamma$ by adding all possible general-valued unary cost functions. Note, $\bar{\Gamma}$ may be different from $\Gamma$ since $\Gamma$ is only guaranteed to have all possible **finite-valued** unary cost functions.

**Proposition 15.** *(a) Graphs $G_\Gamma$ and $G_{\bar{\Gamma}}$ are the same: if $\{(a, b), (a', b')\}$ is a soft (hard) edge in $G_\Gamma$ then it is also a soft (hard) edge in $G_{\bar{\Gamma}}$, and vice versa. (b) If $\bar{\Gamma}$ is NP-hard then so is $\Gamma$.*

A proof of this fact is given in Appendix A. It shows that it suffices to prove theorem 13 for language $\bar{\Gamma}$. Indeed, if $\bar{\Gamma}$ admits an MJN on $\overline{M}$ and a majority polymorphism then so does $\Gamma$. (Note, the definition of $\overline{M}$ is the same for both $\Gamma$ and $\bar{\Gamma}$ by proposition 15(a).) If $\bar{\Gamma}$ does not admit an MJN on $\overline{M}$ or a majority polymorphism then theorem 13 for $\bar{\Gamma}$ and proposition 15(b) would imply that $\Gamma$ is NP-hard.

We can therefore make the following assumption without loss of generality:

**Assumption 1.** $\Gamma$ contains all possible general-valued unary cost functions.

For a language $\Gamma$ let $Feas(\Gamma)$ be the language obtained from $\Gamma$ by converting all finite values of $f$ to 0 for all $f \in \Gamma$, and let $MinHom(\Gamma)$ be the language obtained from $Feas(\Gamma)$ by adding all possible finite-valued unary cost functions. Note, $MinHom(\Gamma)$ corresponds to the *minimum homomorphism* problem [36]. We will need the following fact which is a simple corollary of results of Takhanov [36].

**Theorem 16** ([36]). *If $MinHom(\Gamma)$ does not admit a majority polymorphism then $MinHom(\Gamma)$ is NP-hard.*

Suppose that $\Gamma$ does not admit a majority polymorphism. Clearly, this implies that $MinHom(\Gamma)$ also does not admit a majority polymorphism, and thus is NP-hard by theorem 16. This in turn implies that $\Gamma$ is also NP-hard as can be easily shown (see Appendix B), and so theorem 13 holds in this case. Therefore, we now assume the following:

**Assumption 2.** $\Gamma$ admits a majority polymorphism.

Our last assumption is

**Assumption 3.** $G_\Gamma$ does not have soft self-loops.

Indeed, if it violated then $\Gamma$ is NP-hard by theorem 11(a), so theorem 13 holds.

Our goal will be to show the existence of an MJN multimorphism on $\overline{M}$ under assumptions 1-3. We denote $\langle \sqcap, \sqcup \rangle$ to be an STP multimorphism on $M$ with the properties given in theorem 11(b).

## 4.1 Constructing $\langle \texttt{Mj}_1, \texttt{Mj}_2, \texttt{Mn}_3 \rangle$

Let us introduce function $\mu$ which maps a set $\{a,b,c\} \subseteq D$ with $|\{a,b,c\}| = 3$ to a subset of $\{a,b,c\}$. This subset is defined as follows: $c \in \mu(\{a,b,c\})$ iff there exists binary function $f \in \Gamma^*$ and a pair $(a', b') \in \overline{M}$ such that
$$\text{dom}\, f = \{(a, a'), (b, a'), (c, b')\}$$

**Lemma 17.** *Set $\mu(\{a,b,c\})$ contains at most one label. Furthermore, if $\mu(\{a,b,c\}) = \{c\}$ then $(a,c) \in \overline{M}$ and $(b,c) \in \overline{M}$.*

*Proof.* Suppose that $a, c \in \mu(\{a,b,c\})$ where $a \neq c$, then there exist binary functions $f, g \in \Gamma^*$ and pairs $(a', b'), (a'', b'') \in \overline{M}$ such that
$$\text{dom}\, f = \{(a', a), (b', b), (b', c)\} \qquad \text{dom}\, g = \{(a, a''), (b, a''), (c, b'')\}$$

Consider function
$$h(x', x'') = \min_{x \in D}\{f(x', x) + g(x, x'')\} \tag{4}$$

Clearly, $\text{dom}\, h = \{(a', a''), (b', a''), (b', b'')\}$, so $(a', b') \in \overline{M}$ has an incident soft edge in $G_\Gamma$ - a contradiction.

This second claim of the lemma follows from lemma 9(b). $\square$

For convenience, we define $\mu(\{a,b,c\}) = \varnothing$ if $|\{a,b,c\}| \leq 2$. We are now ready to construct operation $\texttt{MJN} = \langle \texttt{Mj}_1, \texttt{Mj}_2, \texttt{Mn}_3 \rangle$. Given a tuple $(a,b,c) \in D^3$, we define

$$\texttt{MJN}(a,b,c) = \begin{cases} (x, x, y) & \text{if } \{\!\{a,b,c\}\!\} = \{\!\{x,x,y\}\!\}, \{x,y\} \in \overline{M} & (5a) \\ (b \sqcap c, b \sqcup c, a) & \text{if } \mu(\{a,b,c\}) = \{a\} & (5b) \\ (a \sqcap c, a \sqcup c, b) & \text{if } \mu(\{a,b,c\}) = \{b\} & (5c) \\ (a \sqcap b, a \sqcup b, c) & \text{in any other case} & (5d) \end{cases}$$

where $\{\!\{\ldots\}\!\}$ denotes a *multiset*, i.e. elements' multiplicities are taken into account.

**Theorem 18.** *If $f \in \Gamma^*$ and $\boldsymbol{x}, \boldsymbol{y}, \boldsymbol{z} \in \text{dom}\, f$ then*
$$f(\texttt{Mj}_1(\boldsymbol{x},\boldsymbol{y},\boldsymbol{z})) + f(\texttt{Mj}_2(\boldsymbol{x},\boldsymbol{y},\boldsymbol{z})) + f(\texttt{Mn}_3(\boldsymbol{x},\boldsymbol{y},\boldsymbol{z})) \leq f(\boldsymbol{x}) + f(\boldsymbol{y}) + f(\boldsymbol{z}) \tag{6}$$

The remainder of the paper is devoted to the proof of this statement.

## 4.2 Proof of theorem 18: preliminaries

We say that an instance $(f, \boldsymbol{x}, \boldsymbol{y}, \boldsymbol{z})$ is *valid* if $f \in \Gamma^*$ and $\boldsymbol{x}, \boldsymbol{y}, \boldsymbol{z} \in \text{dom} f$. It is *satisfiable* if (6) holds, and *unsatisfiable* otherwise. For a triple $\boldsymbol{x}, \boldsymbol{y}, \boldsymbol{z} \in D^V$ denote $\delta(\boldsymbol{x}, \boldsymbol{y}, \boldsymbol{z}) = \sum_{i \in V} |\{x_i, y_i, z_i\}|$, $\Delta(\boldsymbol{x}, \boldsymbol{y}, \boldsymbol{z}) = \{i \in V \mid x_i \neq y_i\}$ and $\Delta^M(\boldsymbol{x}, \boldsymbol{y}, \boldsymbol{z}) = \{i \in \Delta(\boldsymbol{x}, \boldsymbol{y}, \boldsymbol{z}) \mid \{x_i, y_i, z_i\} = \{a, b\} \in M\}$.

Suppose that an unsatisfiable instance exists. From now on we assume that $(f, \boldsymbol{x}, \boldsymbol{y}, \boldsymbol{z})$ is a lowest unsatisfiable instance with respect to the partial order $\preceq$ defined as the lexicographical order with components

$$(\ \delta(\boldsymbol{x}, \boldsymbol{y}, \boldsymbol{z}), \quad |\Delta(\boldsymbol{x}, \boldsymbol{y}, \boldsymbol{z})|, \quad |\Delta^M(\boldsymbol{x}, \boldsymbol{y}, \boldsymbol{z})|, \quad |\{i \in V \mid \mu(\{x_i, y_i, z_i\}) = \{x_i\}\}|\ ) \tag{7}$$

(the first component is more significant). We denote $\delta_{\min} = \delta(\boldsymbol{x}, \boldsymbol{y}, \boldsymbol{z})$. Thus, we have

**Assumption 4.** *All valid instances $(f, \boldsymbol{x}', \boldsymbol{y}', \boldsymbol{z}')$ with $(\boldsymbol{x}', \boldsymbol{y}', \boldsymbol{z}') \prec (\boldsymbol{x}, \boldsymbol{y}, \boldsymbol{z})$ (and in particular with $\delta(\boldsymbol{x}', \boldsymbol{y}', \boldsymbol{z}') < \delta_{\min}$) are satisfiable, while the instance $(f, \boldsymbol{x}, \boldsymbol{y}, \boldsymbol{z})$ is unsatisfiable.*

We will assume without loss of generality that for any $\boldsymbol{u} \in \text{dom} f$ there holds $u_i \in \{x_i, y_i, z_i\}$ for all $i \in V$. Indeed, this can be achieved by adding unary cost functions $g_i(u_i)$ to $f$ with $\text{dom} g_i = \{x_i, y_i, z_i\}$; this does not affect the satisfiability of $(f, \boldsymbol{x}, \boldsymbol{y}, \boldsymbol{z})$.

The following cases can be easily eliminated:

**Proposition 19.** *The following cases are impossible: (a) $|V| = 1$; (b) $|\{x_i, y_i, z_i\}| = 1$ for some $i \in V$.*

*Proof.* If $|V| = 1$ then (6) is a trivial equality contradicting to the choice of $(f, \boldsymbol{x}, \boldsymbol{y}, \boldsymbol{z})$. Suppose that $x_i = y_i = z_i = a$, $i \in V$. Consider function

$$g(\boldsymbol{u}) = \min_{d \in D} f(d, \boldsymbol{u}) \qquad \forall \boldsymbol{u} \in D^{\hat{V}}$$

where $\hat{V} = V - \{i\}$ and we assumed for simplicity of notation that $i$ corresponds to the first argument of $f$. For an assignment $\boldsymbol{w} \in V$ we denote $\hat{\boldsymbol{w}}$ to be the restriction of $\boldsymbol{w}$ to $\hat{V}$. Clearly, $g \in \Gamma^*$, $g(\hat{\boldsymbol{x}}) = f(\boldsymbol{x})$, $g(\hat{\boldsymbol{y}}) = f(\boldsymbol{y})$, $g(\hat{\boldsymbol{y}}) = f(\boldsymbol{y})$ and $(\hat{\boldsymbol{x}}, \hat{\boldsymbol{y}}, \hat{\boldsymbol{z}}) \prec (\boldsymbol{x}, \boldsymbol{y}, \boldsymbol{z})$, so Assumption 4 gives

$$g(\text{Mj}_1(\hat{\boldsymbol{x}}, \hat{\boldsymbol{y}}, \hat{\boldsymbol{z}})) + g(\text{Mj}_2(\hat{\boldsymbol{x}}, \hat{\boldsymbol{y}}, \hat{\boldsymbol{z}})) + g(\text{Mn}_3(\hat{\boldsymbol{x}}, \hat{\boldsymbol{y}}, \hat{\boldsymbol{z}})) \leq g(\hat{\boldsymbol{x}}) + g(\hat{\boldsymbol{y}}) + g(\hat{\boldsymbol{z}}) = f(\boldsymbol{x}) + f(\boldsymbol{y}) + f(\boldsymbol{z})$$

This implies that $\text{Mj}_1(\hat{\boldsymbol{x}}, \hat{\boldsymbol{y}}, \hat{\boldsymbol{z}}) \in \text{dom} g$ and thus $g(\text{Mj}_1(\hat{\boldsymbol{x}}, \hat{\boldsymbol{y}}, \hat{\boldsymbol{z}})) = f(a, \text{Mj}_1(\hat{\boldsymbol{x}}, \hat{\boldsymbol{y}}, \hat{\boldsymbol{z}})) = f(\text{Mj}_1(\boldsymbol{x}, \boldsymbol{y}, \boldsymbol{z}))$. Similarly, $g(\text{Mj}_2(\hat{\boldsymbol{x}}, \hat{\boldsymbol{y}}, \hat{\boldsymbol{z}})) = f(\text{Mj}_2(\boldsymbol{x}, \boldsymbol{y}, \boldsymbol{z}))$ and $g(\text{Mn}_3(\hat{\boldsymbol{x}}, \hat{\boldsymbol{y}}, \hat{\boldsymbol{z}})) = f(\text{Mn}_3(\boldsymbol{x}, \boldsymbol{y}, \boldsymbol{z}))$, so the inequality above is equivalent to (6). □

It is also easy to show the following fact.

**Proposition 20.** *There exists node $i \in V$ for which operation $\text{MJN}(x_i, y_i, z_i)$ is defined by equation (5a), (5b) or (5c), i.e. either $\{x_i, y_i, z_i\} = \{a, b\} \in \overline{M}$, $\mu(\{x_i, y_i, z_i\}) = \{x_i\}$, or $\mu(\{x_i, y_i, z_i\}) = \{y_i\}$.*

*Proof.* If such a node does not exist then $\text{MJN}(x_i, y_i, z_i)$ is defined by equation (5d) for all nodes $i \in V$, i.e. $\text{MJN}(\boldsymbol{x}, \boldsymbol{y}, \boldsymbol{z}) = (\boldsymbol{x} \sqcap \boldsymbol{y}, \boldsymbol{x} \sqcup \boldsymbol{y}, \boldsymbol{z})$. The fact that $\langle \sqcap, \sqcup \rangle$ is a multimorphism of $f$ then implies inequality (6), contradicting to the choice of $(f, \boldsymbol{x}, \boldsymbol{y}, \boldsymbol{z})$. □

In the next section we show that case (5a) is impossible, while the remaining two cases (5b), (5c) are analysed in section 4.4.

The following equalities are easy to verify; they will be useful for verifying various identities:

$$\alpha \sqcap (\alpha \sqcup \beta) = \alpha \sqcap (\beta \sqcup \alpha) = (\alpha \sqcap \beta) \sqcup \alpha = (\beta \sqcap \alpha) \sqcup \alpha = \alpha \qquad \forall \alpha, \beta \in D \tag{8a}$$

$$\text{MJN}(\alpha, \alpha, \beta) = (\alpha, \alpha, \beta) \qquad \forall \alpha, \beta \in D \tag{8b}$$

$$\{\!\{\text{Mj}_1(\alpha, \beta, \gamma), \text{Mj}_2(\alpha, \beta, \gamma), \text{Mn}_3(\alpha, \beta, \gamma)\}\!\} = \{\!\{\alpha, \beta, \gamma\}\!\} \qquad \forall \alpha, \beta, \gamma \in D \tag{8c}$$

## 4.3 Eliminating case (5a)

We will need the following result.

**Lemma 21.** *Suppose that $i \in V$ is a node with $\{\{x_i, y_i, z_i\}\} = \{\{a, b, b\}\}$ where $\{a, b\} \in \overline{M}$. Let $u \in \{x, y, z\}$ be the labeling with $u_i = a$, and let $u'$ be the labeling obtained from $u$ by setting $u_i = b$. Then $u' \in \text{dom} f$.*

*Proof.* Assume that $u = x$ (the cases $u = y$ and $y = z$ will be entirely analogous). Accordingly, we denote $x' = u'$. By Assumption 2, $f$ admits a majority polymorphism. This implies [1] that $f$ is *decomposable into unary and binary relations*, i.e. there holds

$$u \in \text{dom} f \quad \Leftrightarrow \quad [u_i \in \text{dom} \rho_i \ \forall i \in V \text{ and } (u_i, u_j) \in \text{dom} \rho_{ij} \ \forall i, j \in V, i \neq j]$$

where unary functions $\rho_i \in \Gamma^*$ for $i \in V$ and binary functions $\rho_{ij} \in \Gamma^*$ for distinct $i, j \in V$ are defined as

$$\rho_i(a_i) = \min\{f(u) \mid u_i = a_i\} \qquad \forall a_i \in D$$
$$\rho_{ij}(a_i, a_j) = \min\{f(u) \mid (u_i, u_j) = (a_i, a_j)\} \qquad \forall (a_i, a_j) \in D^2$$

Suppose that $x' \notin \text{dom} f$, then there exists node $j \in V - \{i\}$ such that $(x'_i, x'_j) = (b, x_j) \notin \text{dom} \rho_{ij}$. We must have $(a, x_j), (b, y_j), (b, z_j) \in \text{dom} \rho_{ij}$ since $x, y, z \in \text{dom} f$. This implies, in particular, that $y_j \neq x_j$ and $z_j \neq x_j$. Furthermore, $(a, y_i), (a, z_i) \notin \text{dom} \rho_{ij}$, otherwise pair $(a, b) \in \overline{M}$ would have an incident soft edge in $G_\Gamma$. Denote $a' = x_j$. Two cases are possible:

- $y_j = z_j$. The edge $\{(a, b), (y_j, x_j)\}$ belongs to $G_\Gamma$, therefore $(x_j, y_j) \in \overline{M}$.
- $y_j \neq z_j$. We have $\text{dom} \rho_{ij} = \{(a, x_j), (b, y_j), (b, z_j)\}$, therefore $\mu(\{x_j, y_j, z_j\}) = \{x_j\}$.

In each case $\text{Mj}_1(x_j, y_j, z_j) \neq x_j$, $\text{Mj}_2(x_j, y_j, z_j) \neq x_j$ and $\text{Mn}_3(x_j, y_j, z_j) = x_j$. Now let us "minimize out" variable $x_i$, i.e. define function

$$g(u) = \min_{d \in D} f(d, u) \qquad \forall u \in D^{\hat{V}} \tag{9}$$

where $\hat{V} = V - \{i\}$ and we assumed that $i$ corresponds to the first argument of $f$. For an assignment $u \in V$ we denote $\hat{u}$ to be the restriction of $u$ to $\hat{V}$. Due to the presence of relation $\rho_{ij}$ we have

$$g(\hat{x}) = f(x) \qquad g(\text{Mj}_1(\hat{x}, \hat{y}, \hat{z})) = f(\text{Mj}_1(x, y, z))$$
$$g(\hat{y}) = f(y) \qquad g(\text{Mj}_2(\hat{x}, \hat{y}, \hat{z})) = f(\text{Mj}_2(x, y, z))$$
$$g(\hat{z}) = f(z) \qquad g(\text{Mn}_3(\hat{x}, \hat{y}, \hat{z})) = f(\text{Mn}_3(x, y, z))$$

Since $\delta(\hat{x}, \hat{y}, \hat{z}) < \delta(x, y, z)$, Assumption 4 gives

$$g(\text{Mj}_1(\hat{x}, \hat{y}, \hat{z})) + g(\text{Mj}_2(\hat{x}, \hat{y}, \hat{z})) + g(\text{Mn}_3(\hat{x}, \hat{y}, \hat{z})) \leq f(x) + f(y) + f(z)$$

which is equivalent to (6). □

Let us denote

$$V^M = \{i \in V \mid \{x_i, y_i, z_i\} = \{a, b\} \in M\}$$
$$V^{\overline{M}} = \{i \in V \mid \{x_i, y_i, z_i\} = \{a, b\} \in \overline{M}\}$$
$$V_1^{\overline{M}} = \{i \in V^{\overline{M}} \mid (x_i, y_i, z_i) = (a, b, b)\} \subseteq \Delta(x, y, z)$$
$$V_2^{\overline{M}} = \{i \in V^{\overline{M}} \mid (x_i, y_i, z_i) = (b, a, b)\} \subseteq \Delta(x, y, z)$$
$$V_3^{\overline{M}} = \{i \in V^{\overline{M}} \mid (x_i, y_i, z_i) = (b, b, a)\}$$

We need to show that $V^{\overline{M}}$ is empty.

**Proposition 22.** *Suppose that $i \in V^{\overline{M}}$.*

(a) *If $(x_i, y_i, z_i) = (a, b, b)$ then $\Delta(\boldsymbol{x}, \boldsymbol{y}, \boldsymbol{z}) = \{i\}$ and consequently $V_1^{\overline{M}} = \{i\}$, $\Delta^M(\boldsymbol{x}, \boldsymbol{y}, \boldsymbol{z}) = \varnothing$.*

(b) *If $(x_i, y_i, z_i) = (b, a, b)$ then $\Delta(\boldsymbol{x}, \boldsymbol{y}, \boldsymbol{z}) = \{i\}$ and consequently $V_2^{\overline{M}} = \{i\}$, $\Delta^M(\boldsymbol{x}, \boldsymbol{y}, \boldsymbol{z}) = \varnothing$..*

(c) *If $(x_i, y_i, z_i) = (b, b, a)$ then $V_3^{\overline{M}} = \{i\}$, $|\{x_j, y_j, z_j\}| \le 2$ for all $j \in V$ and $\Delta^M(\boldsymbol{x}, \boldsymbol{y}, \boldsymbol{z}) = \varnothing$.*

*Proof.*

**Part (a)** Suppose that $(x_i, y_i, z_i) = (a, b, b)$ and $\Delta(\boldsymbol{x}, \boldsymbol{y}, \boldsymbol{z})$ is a strict superset of $\{i\}$. Let us define $\boldsymbol{u} = \mathtt{Mn}_3(\boldsymbol{x}, \boldsymbol{y}, \boldsymbol{z})$. It can be checked that $\mathtt{Mj}_1(\boldsymbol{x}, \boldsymbol{x}, \boldsymbol{u}) = \mathtt{Mj}_2(\boldsymbol{x}, \boldsymbol{x}, \boldsymbol{u}) = \boldsymbol{x}$ and $\mathtt{Mn}_3(\boldsymbol{x}, \boldsymbol{x}, \boldsymbol{u}) = \boldsymbol{u}$. Therefore, if we define $\boldsymbol{x}' = \boldsymbol{x}$ and $\boldsymbol{u}' = \boldsymbol{u}$ then the following identities will hold:

$$\begin{aligned}
\mathtt{Mj}_1(\boldsymbol{x}', \boldsymbol{y}, \boldsymbol{z}) &= \mathtt{Mj}_1(\boldsymbol{x}, \boldsymbol{y}, \boldsymbol{z}) & \mathtt{Mj}_1(\boldsymbol{x}, \boldsymbol{x}', \boldsymbol{u}') &= \boldsymbol{x}' \\
\mathtt{Mj}_2(\boldsymbol{x}', \boldsymbol{y}, \boldsymbol{z}) &= \mathtt{Mj}_2(\boldsymbol{x}, \boldsymbol{y}, \boldsymbol{z}) & \mathtt{Mj}_2(\boldsymbol{x}, \boldsymbol{x}', \boldsymbol{u}') &= \boldsymbol{x}' \\
\mathtt{Mn}_3(\boldsymbol{x}', \boldsymbol{y}, \boldsymbol{z}) &= \boldsymbol{u}' & \mathtt{Mn}_3(\boldsymbol{x}, \boldsymbol{x}', \boldsymbol{u}') &= \mathtt{Mn}_3(\boldsymbol{x}, \boldsymbol{y}, \boldsymbol{z})
\end{aligned}$$

Let us modify $\boldsymbol{x}'$ and $\boldsymbol{u}'$ by setting $x_i' = u_i' = b$. It can be checked that the identities above still hold. By lemma 21, $\boldsymbol{x}' \in \mathrm{dom}\, f$. We also have $\delta(\boldsymbol{x}', \boldsymbol{y}, \boldsymbol{z}) < \delta(\boldsymbol{x}, \boldsymbol{y}, \boldsymbol{z})$, so Assumption 4 gives

$$f(\mathtt{Mj}_1(\boldsymbol{x}, \boldsymbol{y}, \boldsymbol{z})) + f(\mathtt{Mj}_2(\boldsymbol{x}, \boldsymbol{y}, \boldsymbol{z})) + f(\boldsymbol{u}') \le f(\boldsymbol{x}') + f(\boldsymbol{y}) + f(\boldsymbol{z}) \tag{10}$$

This implies, in particular, that $\boldsymbol{u}' \in \mathrm{dom}\, f$. We have $(\boldsymbol{x}, \boldsymbol{x}', \boldsymbol{u}') \prec (\boldsymbol{x}, \boldsymbol{y}, \boldsymbol{z})$ since $\Delta(\boldsymbol{x}, \boldsymbol{x}', \boldsymbol{u}') = \{i\}$ and we assumed that $\Delta(\boldsymbol{x}, \boldsymbol{y}, \boldsymbol{z})$ is a strict superset of $\{i\}$. Therefore, Assumption 4 gives

$$f(\boldsymbol{x}') + f(\boldsymbol{x}') + f(\mathtt{Mn}_3(\boldsymbol{x}, \boldsymbol{y}, \boldsymbol{z})) \le f(\boldsymbol{x}) + f(\boldsymbol{x}') + f(\boldsymbol{u}') \tag{11}$$

Summing (10) and (11) gives (6).

**Part (b)** Suppose that $(x_i, y_i, z_i) = (b, a, b)$ and $\Delta(\boldsymbol{x}, \boldsymbol{y}, \boldsymbol{z})$ is a strict subset of $V - \{i\}$. Let $\boldsymbol{u} = \mathtt{Mn}_3(\boldsymbol{x}, \boldsymbol{y}, \boldsymbol{z})$. If we define $\boldsymbol{y}' = \boldsymbol{y}$ and $\boldsymbol{u}' = \boldsymbol{u}$ then the following identities will hold:

$$\begin{aligned}
\mathtt{Mj}_1(\boldsymbol{x}, \boldsymbol{y}', \boldsymbol{z}) &= \mathtt{Mj}_1(\boldsymbol{x}, \boldsymbol{y}, \boldsymbol{z}) & \mathtt{Mj}_1(\boldsymbol{y}, \boldsymbol{y}', \boldsymbol{u}') &= \boldsymbol{y}' \\
\mathtt{Mj}_2(\boldsymbol{x}, \boldsymbol{y}', \boldsymbol{z}) &= \mathtt{Mj}_2(\boldsymbol{x}, \boldsymbol{y}, \boldsymbol{z}) & \mathtt{Mj}_2(\boldsymbol{y}, \boldsymbol{y}', \boldsymbol{u}') &= \boldsymbol{y}' \\
\mathtt{Mn}_3(\boldsymbol{x}, \boldsymbol{y}', \boldsymbol{z}) &= \boldsymbol{u}' & \mathtt{Mn}_3(\boldsymbol{y}, \boldsymbol{y}', \boldsymbol{u}') &= \mathtt{Mn}_3(\boldsymbol{x}, \boldsymbol{y}, \boldsymbol{z})
\end{aligned}$$

Let us modify $\boldsymbol{y}'$ and $\boldsymbol{u}'$ by setting $y_i' = u_i' = b$. It can be checked that the identities above still hold. The rest of the proof is analogous to the proof for part (a).

**Part (c)** Suppose that $(x_i, y_i, z_i) = (b, b, a)$ and (c) does not hold. Let $\boldsymbol{u} = \mathtt{Mn}_3(\boldsymbol{x}, \boldsymbol{y}, \boldsymbol{z})$. If we define $\boldsymbol{z}' = \boldsymbol{z}$ and $\boldsymbol{u}' = \boldsymbol{u}$ then the following identities will hold:

$$\begin{aligned}
\mathtt{Mj}_1(\boldsymbol{x}, \boldsymbol{y}, \boldsymbol{z}') &= \mathtt{Mj}_1(\boldsymbol{x}, \boldsymbol{y}, \boldsymbol{z}) & \mathtt{Mj}_1(\boldsymbol{z}, \boldsymbol{z}', \boldsymbol{u}') &= \boldsymbol{z}' \\
\mathtt{Mj}_2(\boldsymbol{x}, \boldsymbol{y}, \boldsymbol{z}') &= \mathtt{Mj}_2(\boldsymbol{x}, \boldsymbol{y}, \boldsymbol{z}) & \mathtt{Mj}_2(\boldsymbol{z}, \boldsymbol{z}', \boldsymbol{u}') &= \boldsymbol{z}' \\
\mathtt{Mn}_3(\boldsymbol{x}, \boldsymbol{y}, \boldsymbol{z}') &= \boldsymbol{u}' & \mathtt{Mn}_3(\boldsymbol{z}, \boldsymbol{z}', \boldsymbol{u}') &= \mathtt{Mn}_3(\boldsymbol{x}, \boldsymbol{y}, \boldsymbol{z})
\end{aligned}$$

Let us modify $\boldsymbol{z}'$ and $\boldsymbol{u}'$ by setting $z_i' = u_i' = b$. It can be checked that the identities above still hold.

We claim that $(*)$ $(\boldsymbol{z}, \boldsymbol{z}', \boldsymbol{u}') \prec (\boldsymbol{x}, \boldsymbol{y}, \boldsymbol{z})$. Indeed, since (c) does not hold we must have one of the following:

- $V_3^{\overline{M}}$ contains another node $j$ besides $i$. Then $(*)$ holds since $|\{z_j, z_j', u_j'\}| = 1 < |\{x_j, y_j, z_j\}| = 2$.

- $|\{x_j, y_j, z_j\}| = 3$ for some $j \in V$. Then $(*)$ holds since $|\{z_j, z_j', u_j'\}| \le 2$.

- $|\Delta^M(\boldsymbol{x}, \boldsymbol{y}, \boldsymbol{z})| \ge 1$. Then $(*)$ holds since $|\Delta(\boldsymbol{z}, \boldsymbol{z}', \boldsymbol{u}')| = 1 \le |\Delta^M(\boldsymbol{x}, \boldsymbol{y}, \boldsymbol{z})| \le |\Delta(\boldsymbol{x}, \boldsymbol{y}, \boldsymbol{z})|$ and $|\Delta^M(\boldsymbol{z}, \boldsymbol{z}', \boldsymbol{u}')| = 0$.

The rest of the proof is analogous to the proof for part (a). $\square$

Next, we show that if $V^{\overline{M}}$ is non-empty then $V^M$ is empty. By proposition 22 we know that in this case $\Delta^M(\boldsymbol{x}, \boldsymbol{y}, \boldsymbol{z})$ is empty. Thus, if $V^{\overline{M}} \ne \varnothing$ and $i \in V^M$ then we must have $(x_i, y_i, z_i) = (b, b, a)$. This case is eliminated by the following proposition.

**Proposition 23.** *For node $i \in V$ the following situations are impossible:*

S1  $(x_i, y_i, z_i) = (b, b, a)$, $(a, b) \in M$, $a \sqcup b = b$.

S2  $(x_i, y_i, z_i) = (b, b, a)$, $(a, b) \in M$, $a \sqcap b = b$.

*Proof.*
**Case S1** Let us define $\boldsymbol{u} = \mathtt{Mn}_3(\boldsymbol{x}, \boldsymbol{y}, \boldsymbol{z})$. By inspecting each case (5a)-(5d) and using equations (8) one can check that $\boldsymbol{u} \sqcup \boldsymbol{z} = \boldsymbol{z}$ and consequently $\boldsymbol{u} \sqcap \boldsymbol{z} = \boldsymbol{u}$. Therefore, if we define $\boldsymbol{z}' = \boldsymbol{z}$ and $\boldsymbol{u}' = \boldsymbol{u}$ then the following identities will hold:

$$\begin{array}{rclrcl}
\mathtt{Mj}_1(\boldsymbol{x}, \boldsymbol{y}, \boldsymbol{z}') &=& \mathtt{Mj}_1(\boldsymbol{x}, \boldsymbol{y}, \boldsymbol{z}) & \boldsymbol{u}' \sqcap \boldsymbol{z} &=& \mathtt{Mn}_3(\boldsymbol{x}, \boldsymbol{y}, \boldsymbol{z}) \\
\mathtt{Mj}_2(\boldsymbol{x}, \boldsymbol{y}, \boldsymbol{z}') &=& \mathtt{Mj}_2(\boldsymbol{x}, \boldsymbol{y}, \boldsymbol{z}) & \boldsymbol{u}' \sqcup \boldsymbol{z} &=& \boldsymbol{z}' \\
\mathtt{Mn}_3(\boldsymbol{x}, \boldsymbol{y}, \boldsymbol{z}') &=& \boldsymbol{u}' & &&
\end{array}$$

Let us modify $\boldsymbol{z}'$ and $\boldsymbol{u}'$ by setting $z_i' = u_i' = b$, so that we have

$$\begin{array}{ll}
\text{-}\ a = z_i = u_i\ = \mathtt{Mn}_3(x_i, y_i, z_i) & \\
\text{-}\ b = z_i' = u_i'\ = \mathtt{Mj}_{1,2}(x_i, y_i, z_i) \quad (= x_i = y_i) & (a \sqcup b = b)
\end{array}$$

It can be checked that the identities above still hold. We have $\delta(\boldsymbol{x}, \boldsymbol{y}, \boldsymbol{z}') < \delta(\boldsymbol{x}, \boldsymbol{y}, \boldsymbol{z})$, so Assumption 4 gives

$$f(\mathtt{Mj}_1(\boldsymbol{x}, \boldsymbol{y}, \boldsymbol{z})) + f(\mathtt{Mj}_2(\boldsymbol{x}, \boldsymbol{y}, \boldsymbol{z})) + f(\boldsymbol{u}') \ \le\ f(\boldsymbol{x}) + f(\boldsymbol{y}) + f(\boldsymbol{z}') \tag{12}$$

assuming that $\boldsymbol{z}' \in \mathrm{dom}\, f$, and the fact that $\langle \sqcap, \sqcup \rangle$ is a multimorphism of $f$ gives

$$f(\mathtt{Mn}_3(\boldsymbol{x}, \boldsymbol{y}, \boldsymbol{z})) + f(\boldsymbol{z}') \ \le\ f(\boldsymbol{u}') + f(\boldsymbol{z}) \tag{13}$$

assuming that $\boldsymbol{u}' \in \mathrm{dom}\, f$. If $\boldsymbol{z}' \in \mathrm{dom}\, f$ then (12) implies that $\boldsymbol{u}' \in \mathrm{dom}\, f$; summing (12) and (13) gives (6). We thus assume that $\boldsymbol{z}' \notin \mathrm{dom}\, f$, then (13) implies that $\boldsymbol{u}' \notin \mathrm{dom}\, f$.

Let $C$ be a sufficiently large constant, namely $C > f(\boldsymbol{x}) + f(\boldsymbol{y}) + f(\boldsymbol{z})$. Consider function

$$g(\boldsymbol{u}) = \min_{d \in D}\{[d = a] \cdot C + f(d, \boldsymbol{u})\} \qquad \forall \boldsymbol{u} \in D^{\hat{V}} \tag{14}$$

where $\hat{V} = V - \{i\}$, $[\cdot]$ is the Iverson bracket (it is 1 if its argument is true, and 0 otherwise) and we assumed for simplicity of notation that $i$ corresponds to the first argument of $f$. For an assignment $\boldsymbol{w} \in V$ we denote $\hat{\boldsymbol{w}}$ to be the restriction of $\boldsymbol{w}$ to $\hat{V}$. We can write

$$g(\hat{\boldsymbol{z}}) = f(\boldsymbol{z}) + C \qquad g(\hat{\boldsymbol{x}}) = f(\boldsymbol{x}) \qquad g(\hat{\boldsymbol{y}}) = f(\boldsymbol{y}) \qquad g(\hat{\boldsymbol{u}}) = f(\boldsymbol{u}) + C$$

where the first equation holds since $(b, \hat{\boldsymbol{z}}) = \boldsymbol{z}' \notin \operatorname{dom} f$ and the last equation holds since $(b, \hat{\boldsymbol{u}}) = \boldsymbol{u}' \notin \operatorname{dom} f$. Assumption 4 gives

$$g(\mathtt{Mj}_1(\hat{\boldsymbol{x}}, \hat{\boldsymbol{y}}, \hat{\boldsymbol{z}})) + g(\mathtt{Mj}_2(\hat{\boldsymbol{x}}, \hat{\boldsymbol{y}}, \hat{\boldsymbol{z}})) + g(\mathtt{Mn}_3(\hat{\boldsymbol{x}}, \hat{\boldsymbol{y}}, \hat{\boldsymbol{z}})) \leq g(\hat{\boldsymbol{x}}) + g(\hat{\boldsymbol{y}}) + g(\hat{\boldsymbol{z}})$$
$$g(\mathtt{Mj}_1(\hat{\boldsymbol{x}}, \hat{\boldsymbol{y}}, \hat{\boldsymbol{z}})) + g(\mathtt{Mn}_3(\hat{\boldsymbol{x}}, \hat{\boldsymbol{y}}, \hat{\boldsymbol{z}})) + [(f(\boldsymbol{u}) + C] \leq f(\boldsymbol{x}) + f(\boldsymbol{y}) + [f(\boldsymbol{z}) + C]$$

Therefore, $g(\mathtt{Mj}_1(\hat{\boldsymbol{x}}, \hat{\boldsymbol{y}}, \hat{\boldsymbol{z}})) < C$, and thus $g(\mathtt{Mj}_1(\hat{\boldsymbol{x}}, \hat{\boldsymbol{y}}, \hat{\boldsymbol{z}})) = f(b, \mathtt{Mj}_1(\hat{\boldsymbol{x}}, \hat{\boldsymbol{y}}, \hat{\boldsymbol{z}})) = f(\mathtt{Mj}_1(\boldsymbol{x}, \boldsymbol{y}, \boldsymbol{z}))$. Similarly, $g(\mathtt{Mj}_2(\hat{\boldsymbol{x}}, \hat{\boldsymbol{y}}, \hat{\boldsymbol{z}})) = f(b, \mathtt{Mj}_2(\hat{\boldsymbol{x}}, \hat{\boldsymbol{y}}, \hat{\boldsymbol{z}})) = f(\mathtt{Mj}_2(\boldsymbol{x}, \boldsymbol{y}, \boldsymbol{z}))$, and hence the inequality above is equivalent to (6).

**Case S2** Let us define $\boldsymbol{u} = \mathtt{Mn}_3(\boldsymbol{x}, \boldsymbol{y}, \boldsymbol{z})$. It can be checked that $\boldsymbol{z} \sqcap \boldsymbol{u} = \boldsymbol{z}$ and consequently $\boldsymbol{z} \sqcup \boldsymbol{u} = \boldsymbol{u}$. Therefore, if we define $\boldsymbol{z}' = \boldsymbol{z}$ and $\boldsymbol{u}' = \boldsymbol{u}$ then the following identities will hold:

$$\mathtt{Mj}_1(\boldsymbol{x}, \boldsymbol{y}, \boldsymbol{z}') = \mathtt{Mj}_1(\boldsymbol{x}, \boldsymbol{y}, \boldsymbol{z}) \qquad \boldsymbol{z} \sqcap \boldsymbol{u}' = \boldsymbol{z}'$$
$$\mathtt{Mj}_2(\boldsymbol{x}, \boldsymbol{y}, \boldsymbol{z}') = \mathtt{Mj}_2(\boldsymbol{x}, \boldsymbol{y}, \boldsymbol{z}) \qquad \boldsymbol{z} \sqcup \boldsymbol{u}' = \mathtt{Mn}_3(\boldsymbol{x}, \boldsymbol{y}, \boldsymbol{z})$$
$$\mathtt{Mn}_3(\boldsymbol{x}, \boldsymbol{y}, \boldsymbol{z}') = \boldsymbol{u}'$$

Let us modify $\boldsymbol{z}'$ and $\boldsymbol{u}'$ by setting $z'_i = u'_i = b$, so that we have

- $a = z_i = u_i = \mathtt{Mn}_3(x_i, y_i, z_i)$
- $b = z'_i = u'_i = \mathtt{Mj}_{1,2}(x_i, y_i, z_i) \quad (= x_i = y_i)$ $\qquad (a \sqcap b = b)$

It can be checked that the identities above still hold. The rest of the proof proceeds analogously to the proof for the case S1. □

We are now ready to prove the following fact.

**Proposition 24.** *Set $V^{\overline{M}}$ is empty.*

*Proof.* Suppose that $V^{\overline{M}} \neq \emptyset$. As we just showed, we must have $V^M = \emptyset$. For each $i \in V$ we also have $|\{x_i, y_i, z_i\}| \neq 1$ by proposition 19 and $|\{x_i, y_i, z_i\}| \neq 3$ by proposition 22. Therefore, $V = V^{\overline{M}}$. Proposition 22 implies that each of the sets $V_1^{\overline{M}}$, $V_2^{\overline{M}}$, $V_3^{\overline{M}}$ contains at most one node, and furthermore $|V_1^{\overline{M}} \cup V_2^{\overline{M}}| \leq 1$. Since $|V| \geq 2$ by proposition 19, we conclude that $V = \{i, j\}$ where $i \in V_3^{\overline{M}}$ and $j \in V_1^{\overline{M}} \cup V_2^{\overline{M}}$.

Suppose that $j \in V_1^{\overline{M}}$, then we have $\boldsymbol{x} = (b, a')$, $\boldsymbol{y} = (b, b')$, $\boldsymbol{z} = (a, b')$ where $\{a, b\}, \{a', b'\} \in \overline{M}$. Inequality (6) reduces to

$$f(b, b') + f(b, b') + f(a, a') \leq f(b, a') + f(b, b') + f(a, b') \tag{15}$$

We must have $f(a, a') + f(b, b') = f(a, b') + f(b, a')$, otherwise $(a, b)$ would have a soft incident edge in $G_\Gamma$ contradicting to lemma 9(c). Therefore, (15) is an equality. The case $j \in V_2^{\overline{M}}$ is completely analogous. Proposition 24 is proved. □

## 4.4 Eliminating cases (5b) and (5c)

Propositions 20 and 24 show that there must exist node $i \in V$ with $\mu(\{x_i, y_i, z_i\}) = \{x_i\}$ or $\mu(\{x_i, y_i, z_i\}) = \{y_i\}$. In this section we show that this leads to a contradiction, thus proving theorem 18.

Consider variable $i \in V$ with $\mu(\{x_i, y_i, z_i\}) = \{a\} \neq \varnothing$. We say that another variable $j \in V - \{i\}$ is a *control variable* for $i$ if $\{x_j, y_j, z_j\} = \{\alpha, \beta\} \in \overline{M}$ and for any labeling $\boldsymbol{u} \in \text{dom} f$ the following is true: $u_i = a$ iff $u_j = \alpha$. This implies the following property:

**Proposition 25.** *Suppose that variable $i \in V$ with $\mu(\{x_i, y_i, z_i\}) = \{a\}$ has a control variable. Let $\boldsymbol{u}$, $\boldsymbol{v}$, $\boldsymbol{w}$ be a permutation of $\boldsymbol{x}, \boldsymbol{y}, \boldsymbol{z}$ such that $u_i = a$. Then*

- *Any labeling obtained from one of the labelings in $\{\boldsymbol{u}, \text{Mn}_3(\boldsymbol{x}, \boldsymbol{y}, \boldsymbol{z})\}$ by changing the label of $i$ from $a$ to $v_i$ or $w_i$ does not belong to $\text{dom} f$.*

- *Any labeling obtained from one of the labelings in $\{\boldsymbol{v}, \boldsymbol{w}, \text{Mj}_1(\boldsymbol{x}, \boldsymbol{y}, \boldsymbol{z}), \text{Mj}_2(\boldsymbol{x}, \boldsymbol{y}, \boldsymbol{z})\}$ by changing the label of $i$ from $\{v_i, w_i\}$ to $a$ does not belong to $\text{dom} f$.*

Let $(f, \boldsymbol{x}, \boldsymbol{y}, \boldsymbol{z})$ be a valid instance and $i \in V$ be a variable with $\mu(\{x_i, y_i, z_i\}) \neq \varnothing$. If $i$ does not have a control variable then we can define another valid instance $(\bar{f}, \bar{\boldsymbol{x}}, \bar{\boldsymbol{y}}, \bar{\boldsymbol{z}})$ with the set of variables $\bar{V} = V \cup \{j\}$, $j \neq V$ as follows:
$$\bar{f}(\boldsymbol{u}) = f(\hat{\boldsymbol{u}}) + g(u_i, u_j) \qquad \forall \boldsymbol{u} \in D^{\bar{V}}$$
where $g$ is a binary function taken from the definition of the set $\mu(\{x_i, y_i, z_i\})$ and $\hat{\boldsymbol{u}}$ is the restriction of $\boldsymbol{u}$ to $V$. Labelings $\bar{\boldsymbol{x}}, \bar{\boldsymbol{y}}, \bar{\boldsymbol{z}}$ are obtained by extending $\boldsymbol{x}, \boldsymbol{y}, \boldsymbol{z}$ to $\bar{V}$ in the unique way so that $(\bar{f}, \bar{\boldsymbol{x}}, \bar{\boldsymbol{y}}, \bar{\boldsymbol{z}})$ is a valid instance. Clearly, in the new instance variable $i$ does have a control variable. Furthermore, this transformation does not affect the satisfiability of the instance, and $\delta(\boldsymbol{x}, \boldsymbol{y}, \boldsymbol{z})$ is increased by 2. Such transformation will be used below; after introducing control variable $j$ we will "minimize out" variable $x_i$, which will decrease $\delta(\boldsymbol{x}, \boldsymbol{y}, \boldsymbol{z})$ by 3.

If $\mu(\{a, b, c\}) = \{c\}$ then we will illustrate this fact using the following diagram:

**Proposition 26.** *For node $i \in V$ the following situations are impossible:*

T1 $\mu(\{x_i, y_i, z_i\}) = \{y_i\}$, $(x_i, z_i) \in M$, $x_i \sqcap z_i = z_i$.

T2 $\mu(\{x_i, y_i, z_i\}) = \{y_i\}$, $(x_i, z_i) \in M$, $x_i \sqcup z_i = z_i$.

T3 $\mu(\{x_i, y_i, z_i\}) = \{x_i\}$, $(y_i, z_i) \in M$, $y_i \sqcup z_i = z_i$.

T4 $\mu(\{x_i, y_i, z_i\}) = \{x_i\}$, $(y_i, z_i) \in M$, $y_i \sqcap z_i = z_i$.

*Proof.* We will analyse cases T1-T4 separately, and will derive a contradiction in each case.

**Case T1** Let us define $\boldsymbol{u} = \text{Mj}_2(\boldsymbol{x}, \boldsymbol{y}, \boldsymbol{z})$. It can be checked that $\boldsymbol{x} \sqcap \boldsymbol{u} = \boldsymbol{x}$ and consequently $\boldsymbol{x} \sqcup \boldsymbol{u} = \boldsymbol{u}$. Therefore, if we define $\boldsymbol{x}' = \boldsymbol{x}$ and $\boldsymbol{u}' = \boldsymbol{u}$ then the following identities will hold:

$$\begin{aligned}
\text{Mj}_1(\boldsymbol{x}', \boldsymbol{y}, \boldsymbol{z}) &= \text{Mj}_1(\boldsymbol{x}, \boldsymbol{y}, \boldsymbol{z}) & \boldsymbol{x} \sqcap \boldsymbol{u}' &= \boldsymbol{x}' \\
\text{Mj}_2(\boldsymbol{x}', \boldsymbol{y}, \boldsymbol{z}) &= \boldsymbol{u}' & \boldsymbol{x} \sqcup \boldsymbol{u}' &= \text{Mj}_2(\boldsymbol{x}, \boldsymbol{y}, \boldsymbol{z}) = \boldsymbol{u} \qquad (16) \\
\text{Mn}_3(\boldsymbol{x}', \boldsymbol{y}, \boldsymbol{z}) &= \text{Mn}_3(\boldsymbol{x}, \boldsymbol{y}, \boldsymbol{z})
\end{aligned}$$

Let us modify $\boldsymbol{x}', \boldsymbol{u}'$ by setting $x'_i = u'_i = \text{Mj}_1(x_i, y_i, z_i)$ so that we have

$$\begin{aligned}
a &= x_i = u_i &&= \text{Mj}_2(x_i, y_i, z_i) \\
b &= x'_i = u'_i &&= \text{Mj}_1(x_i, y_i, z_i) \ (= z_i) &&\quad (a \sqcap b = b) \\
c &&&= \text{Mn}_3(x_i, y_i, z_i) \ (= y_i)
\end{aligned}$$

where we denoted $(a, b, c) = (x_i, z_i, y_i)$. It can be checked that identities (16) still hold, and furthermore $\delta(\boldsymbol{x}', \boldsymbol{y}, \boldsymbol{z}) < \delta(\boldsymbol{x}, \boldsymbol{y}, \boldsymbol{z})$. Assumption 4 gives

$$f(\text{Mj}_1(\boldsymbol{x}, \boldsymbol{y}, \boldsymbol{z})) + f(\boldsymbol{u}') + f(\text{Mn}_3(\boldsymbol{x}, \boldsymbol{y}, \boldsymbol{z})) \ \leq \ f(\boldsymbol{x}') + f(\boldsymbol{y}) + f(\boldsymbol{z}) \tag{17}$$

assuming that $\boldsymbol{x}' \in \text{dom} f$, and the fact that $\langle \sqcap, \sqcup \rangle$ is a multimorphism of $f$ gives

$$f(\boldsymbol{x}') + f(\text{Mj}_2(\boldsymbol{x}, \boldsymbol{y}, \boldsymbol{z})) \ \leq \ f(\boldsymbol{x}) + f(\boldsymbol{u}') \tag{18}$$

assuming that $\boldsymbol{u}' \in \text{dom} f$. If $\boldsymbol{x}' \in \text{dom} f$ then (17) implies that $\boldsymbol{u}' \in \text{dom} f$; summing (17) and (18) gives (6). We thus assume that $\boldsymbol{x}' \notin \text{dom} f$, then (18) implies that $\boldsymbol{u}' \notin \text{dom} f$.

Let us add a control variable for $i$ using the transformation described above. For simplicity, we do not change the notation, so we assume that $V$ now contains a control variable for $i$ and $\boldsymbol{x}, \boldsymbol{y}, \boldsymbol{z}, \boldsymbol{u}, \boldsymbol{x}', \boldsymbol{u}'$ have been extended to the new set accordingly. We have $\delta(\boldsymbol{x}, \boldsymbol{y}, \boldsymbol{z}) = \delta_{\min} + 2$.

Let $C$ be a sufficiently large constant, namely $C > f(\boldsymbol{x}) + f(\boldsymbol{y}) + f(\boldsymbol{z})$. Consider function

$$g(\boldsymbol{u}) = \min_{d \in D} \{[d = a] \cdot C + f(d, \boldsymbol{u})\} \qquad \forall \boldsymbol{u} \in D^{\hat{V}} \tag{19}$$

where $\hat{V} = V - \{i\}$, $[\cdot]$ is the Iverson bracket (it returns 1 if its argument is true and 0 otherwise) and we assumed for simplicity of notation that $i$ corresponds to the first argument of $f$. For an assignment $\boldsymbol{w} \in V$ we denote $\hat{\boldsymbol{w}}$ to be the restriction of $\boldsymbol{w}$ to $\hat{V}$. We can write

$$g(\hat{\boldsymbol{x}}) = f(\boldsymbol{x}) + C \qquad g(\hat{\boldsymbol{y}}) = f(\boldsymbol{y}) \qquad g(\hat{\boldsymbol{z}}) = f(\boldsymbol{z}) \qquad g(\hat{\boldsymbol{u}}) = f(\boldsymbol{u}) + C \tag{20}$$

To show the first equation, observe that the the minimum in (19) cannot be achieved at $d = b$ since $(b, \hat{\boldsymbol{x}}) = \boldsymbol{x}' \notin \text{dom} f$, and also the minimum cannot be achieved at $d = c$ by proposition 25. Therefore, $g(\hat{\boldsymbol{x}}) = g(a, \hat{\boldsymbol{x}}) = f(\boldsymbol{x}) + C$. Other equations can be derived similarly.

Clearly, $(g, \hat{\boldsymbol{x}}, \hat{\boldsymbol{y}}, \hat{\boldsymbol{z}})$ is a valid instance and $\delta(\hat{\boldsymbol{x}}, \hat{\boldsymbol{y}}, \hat{\boldsymbol{z}}) = \delta_{\min} - 1$, so Assumption 4 gives

$$\begin{aligned}
g(\text{Mj}_1(\hat{\boldsymbol{x}}, \hat{\boldsymbol{y}}, \hat{\boldsymbol{z}})) + g(\text{Mj}_2(\hat{\boldsymbol{x}}, \hat{\boldsymbol{y}}, \hat{\boldsymbol{z}})) + g(\text{Mn}_3(\hat{\boldsymbol{x}}, \hat{\boldsymbol{y}}, \hat{\boldsymbol{z}})) &\leq g(\hat{\boldsymbol{x}}) + g(\hat{\boldsymbol{y}}) + g(\hat{\boldsymbol{z}}) \\
g(\text{Mj}_1(\hat{\boldsymbol{x}}, \hat{\boldsymbol{y}}, \hat{\boldsymbol{z}})) + [f(\boldsymbol{u}) + C] + g(\text{Mn}_3(\hat{\boldsymbol{x}}, \hat{\boldsymbol{y}}, \hat{\boldsymbol{z}})) &\leq [f(\boldsymbol{x}) + C] + f(\boldsymbol{y}) + f(\boldsymbol{z})
\end{aligned}$$

Therefore, $g(\text{Mj}_1(\hat{\boldsymbol{x}}, \hat{\boldsymbol{y}}, \hat{\boldsymbol{z}})) < C$, and thus $g(\text{Mj}_1(\hat{\boldsymbol{x}}, \hat{\boldsymbol{y}}, \hat{\boldsymbol{z}})) = f(b, \text{Mj}_1(\hat{\boldsymbol{x}}, \hat{\boldsymbol{y}}, \hat{\boldsymbol{z}})) = f(\text{Mj}_1(\boldsymbol{x}, \boldsymbol{y}, \boldsymbol{z}))$. (Note, labeling $(c, \text{Mj}_1(\hat{\boldsymbol{x}}, \hat{\boldsymbol{y}}, \hat{\boldsymbol{z}}))$ is not in $\text{dom} f$ by proposition 25.) Similarly, $g(\text{Mn}_3(\hat{\boldsymbol{x}}, \hat{\boldsymbol{y}}, \hat{\boldsymbol{z}})) = f(c, \text{Mn}_3(\hat{\boldsymbol{x}}, \hat{\boldsymbol{y}}, \hat{\boldsymbol{z}})) = f(\text{Mn}_3(\boldsymbol{x}, \boldsymbol{y}, \boldsymbol{z}))$, and hence the inequality above is equivalent to (6).

**Case** T2  Let us define $\boldsymbol{u} = \text{Mj}_1(\boldsymbol{x}, \boldsymbol{y}, \boldsymbol{z})$. It can be checked that $\boldsymbol{u} \sqcup \boldsymbol{x} = \boldsymbol{x}$ and consequently $\boldsymbol{u} \sqcap \boldsymbol{x} = \boldsymbol{u}$. Therefore, if we define $\boldsymbol{x}' = \boldsymbol{x}$ and $\boldsymbol{u}' = \boldsymbol{u}$ then the following identities will hold:

$$\begin{aligned}
\text{Mj}_1(\boldsymbol{x}', \boldsymbol{y}, \boldsymbol{z}) &= \boldsymbol{u}' & \boldsymbol{u}' \sqcap \boldsymbol{x} &= \text{Mj}_1(\boldsymbol{x}, \boldsymbol{y}, \boldsymbol{z}) = \boldsymbol{u} \\
\text{Mj}_2(\boldsymbol{x}', \boldsymbol{y}, \boldsymbol{z}) &= \text{Mj}_2(\boldsymbol{x}, \boldsymbol{y}, \boldsymbol{z}) & \boldsymbol{u}' \sqcup \boldsymbol{x} &= \boldsymbol{x}' \\
\text{Mn}_3(\boldsymbol{x}', \boldsymbol{y}, \boldsymbol{z}) &= \text{Mn}_3(\boldsymbol{x}, \boldsymbol{y}, \boldsymbol{z})
\end{aligned}$$

Let us modify $\boldsymbol{x}', \boldsymbol{u}'$ by setting $x'_i = u'_i = \mathtt{Mj}_2(x_i, y_i, z_i)$ so that we have

$$\begin{cases} a = x_i = u_i & = \mathtt{Mj}_1(x_i, y_i, z_i) \\ b = x'_i = u'_i & = \mathtt{Mj}_2(x_i, y_i, z_i) \ (= z_i) \\ c & = \mathtt{Mn}_3(x_i, y_i, z_i) \ (= y_i) \end{cases} \qquad (a \sqcup b = b)$$

It can be checked that the identities above still hold. The rest of the proof proceeds analogously to the proof for the case T1.

**Case T3**  Let us define $\boldsymbol{u} = \mathtt{Mj}_1(\boldsymbol{x}, \boldsymbol{y}, \boldsymbol{z})$. It can be checked that $\boldsymbol{u} \sqcup \boldsymbol{y} = \boldsymbol{y}$ and consequently $\boldsymbol{u} \sqcap \boldsymbol{y} = \boldsymbol{u}$. Therefore, if we define $\boldsymbol{y}' = \boldsymbol{y}$ and $\boldsymbol{u}' = \boldsymbol{u}$ then the following identities will hold:

$$\mathtt{Mj}_1(\boldsymbol{x}, \boldsymbol{y}', \boldsymbol{z}) = \boldsymbol{u}' \qquad\qquad \boldsymbol{u}' \sqcap \boldsymbol{y} = \mathtt{Mj}_1(\boldsymbol{x}, \boldsymbol{y}, \boldsymbol{z}) = \boldsymbol{u}$$
$$\mathtt{Mj}_2(\boldsymbol{x}, \boldsymbol{y}', \boldsymbol{z}) = \mathtt{Mj}_2(\boldsymbol{x}, \boldsymbol{y}, \boldsymbol{z}) \qquad\qquad \boldsymbol{u}' \sqcup \boldsymbol{y} = \boldsymbol{y}'$$
$$\mathtt{Mn}_3(\boldsymbol{x}, \boldsymbol{y}', \boldsymbol{z}) = \mathtt{Mn}_3(\boldsymbol{x}, \boldsymbol{y}, \boldsymbol{z})$$

Let us modify $\boldsymbol{y}', \boldsymbol{u}'$ by setting $y'_i = u'_i = \mathtt{Mj}_2(x_i, y_i, z_i)$ so that we have

$$\begin{cases} a = y_i = u_i & = \mathtt{Mj}_1(x_i, y_i, z_i) \\ b = y'_i = u'_i & = \mathtt{Mj}_2(x_i, y_i, z_i) \ (= z_i) \\ c & = \mathtt{Mn}_3(x_i, y_i, z_i) \ (= x_i) \end{cases} \qquad (a \sqcup b = b)$$

It can be checked that the identities above still hold. The rest of the proof proceeds analogously to the proof for the case T1.

**Case T4**  Let us define $\boldsymbol{u} = \mathtt{Mj}_2(\boldsymbol{x}, \boldsymbol{y}, \boldsymbol{z})$. It can be checked that $\boldsymbol{y} \sqcap \boldsymbol{u} = \boldsymbol{y}$ and consequently $\boldsymbol{y} \sqcup \boldsymbol{u} = \boldsymbol{u}$. Therefore, if we define $\boldsymbol{y}' = \boldsymbol{y}$ and $\boldsymbol{u}' = \boldsymbol{u}$ then the following identities will hold:

$$\mathtt{Mj}_1(\boldsymbol{x}, \boldsymbol{y}', \boldsymbol{z}) = \mathtt{Mj}_1(\boldsymbol{x}, \boldsymbol{y}, \boldsymbol{z}) \qquad\qquad \boldsymbol{y} \sqcap \boldsymbol{u}' = \boldsymbol{y}'$$
$$\mathtt{Mj}_2(\boldsymbol{x}, \boldsymbol{y}', \boldsymbol{z}) = \boldsymbol{u}' \qquad\qquad \boldsymbol{y} \sqcup \boldsymbol{u}' = \mathtt{Mj}_2(\boldsymbol{x}, \boldsymbol{y}, \boldsymbol{z}) = \boldsymbol{u}$$
$$\mathtt{Mn}_3(\boldsymbol{x}, \boldsymbol{y}', \boldsymbol{z}) = \mathtt{Mn}_3(\boldsymbol{x}, \boldsymbol{y}, \boldsymbol{z})$$

Let us modify $\boldsymbol{y}', \boldsymbol{u}'$ by setting $y'_i = u'_i = \mathtt{Mj}_1(x_i, y_i, z_i)$ so that we have

$$\begin{cases} a = y_i = u_i & = \mathtt{Mj}_2(x_i, y_i, z_i) \\ b = y'_i = u'_i & = \mathtt{Mj}_1(x_i, y_i, z_i) \ (= z_i) \\ c & = \mathtt{Mn}_3(x_i, y_i, z_i) \ (= x_i) \end{cases} \qquad (a \sqcap b = b)$$

It can be checked that the identities above still hold. The rest of the proof proceeds analogously to the proof for the case T1. □

There are two possible cases remaining: $\mu(\{x_i, y_i, z_i\}) = \{y_i\}$, $\{x_i, z_i\} \in \overline{M}$ or $\mu(\{x_i, y_i, z_i\}) = \{x_i\}$, $\{y_i, z_i\} \in \overline{M}$. They are eliminated by the next two propositions; we use a slightly different argument.

**Proposition 27.** *For node $i \in V$ the following situation is impossible:*

T5  $\mu(\{x_i, y_i, z_i\}) = \{y_i\}$, $\{x_i, z_i\} \in \overline{M}$.

*Proof.* For a labeling $w \in D^V$ let $\hat{w}$ be the restriction of $w$ to $V - \{i\}$. Two cases are possible.

**Case 1** $(\texttt{Mj}_2(\hat{x}, \hat{y}, \hat{z}), \hat{y}, \hat{z}) \prec (\hat{x}, \hat{y}, \hat{z})$. Let us define $u = \texttt{Mj}_2(x, y, z)$ and $v = \texttt{Mj}_2(u, y, z)$. It can be checked that $\texttt{MJN}(u, v, z) = (u, v, z)$. [1] Therefore, if we define $z' = z$ and $u' = u$ then the following identities will hold:

$$\texttt{Mj}_1(x, y, z') = \texttt{Mj}_1(x, y, z) \qquad \texttt{Mj}_1(u', v, z) = \texttt{Mj}_2(x, y, z) = u \qquad v = \texttt{Mj}_2(u', y, z)$$
$$\texttt{Mj}_2(x, y, z') = u' \qquad \texttt{Mj}_2(u', v, z) = v$$
$$\texttt{Mn}_3(x, y, z') = \texttt{Mn}_3(x, y, z) \qquad \texttt{Mn}_3(u', v, z) = z'$$

Let us modify $z'$ and $u'$ according to the following diagram:

$$\begin{array}{l} a = z_i = u_i = \texttt{Mj}_2(x_i, y_i, z_i) \ (= v_i) \\ b = z'_i = u'_i = \texttt{Mj}_1(x_i, y_i, z_i) \ (= x_i) \\ c = \phantom{z'_i = u'_i} = \texttt{Mn}_3(x_i, y_i, z_i) \ (= y_i) \end{array}$$

It can be checked that the identities above still hold. The assumption of Case 1 gives $(u', y, z) \prec (x, y, z)$ (note that $u'_i = x_i$). Therefore, the fact that $v = \texttt{Mj}_2(u', y, z)$ and Assumption 4 give the following relationship: $(*)$ if $u' \in \text{dom} f$ then $v \in \text{dom} f$.

We have $\delta(x, y, z') < \delta(x, y, z)$ and $\delta(u', v, z) < \delta(x, y, z)$, so Assumption 4 gives

$$f(\texttt{Mj}_1(x, y, z)) + f(u') + f(\texttt{Mn}_3(x, y, z)) \leq f(x) + f(y) + f(z') \tag{21}$$

assuming that $z' \in \text{dom} f$, and

$$f(\texttt{Mj}_2(x, y, z)) + f(v) + f(z') \leq f(u') + f(v) + f(z) \tag{22}$$

assuming that $u', v \in \text{dom} f$. If $z' \in \text{dom} f$ then (21) implies that $u' \in \text{dom} f$, and so $(*)$ implies that $v \in \text{dom} f$. Summing (21) and (22) gives (6). We thus assume that $z' \notin \text{dom} f$, then we have $u' \notin \text{dom} f$. (If $u' \in \text{dom} f$ then $(*)$ gives $v \in \text{dom} f$, and equation (22) then gives $z' \in \text{dom} f$ - a contradiction.)

The rest of the argument proceeds similar to that for the case T1. Let us add a control variable for $i$ (again, without changing the notation). Consider function

$$g(u) = \min_{d \in D} \{[d = a] \cdot C + f(d, u)\} \qquad \forall u \in D^{\hat{V}}$$

where $\hat{V} = V - \{i\}$ and $C > f(x) + f(y) + f(z)$ is a sufficiently large constant. We can write

$$g(\hat{z}) = f(z) + C \qquad g(\hat{x}) = f(x) \qquad g(\hat{y}) = f(y) \qquad g(\hat{u}) = f(u) + C$$

Clearly, $(g, \hat{x}, \hat{y}, \hat{z})$ is a valid instance and $\delta(\hat{x}, \hat{y}, \hat{z}) = \delta_{\min} - 1$, so Assumption 4 gives

$$g(\texttt{Mj}_1(\hat{x}, \hat{y}, \hat{z})) + g(\texttt{Mj}_2(\hat{x}, \hat{y}, \hat{z})) + g(\texttt{Mn}_3(\hat{x}, \hat{y}, \hat{z})) \leq g(\hat{x}) + g(\hat{y}) + g(\hat{z})$$
$$g(\texttt{Mj}_1(\hat{x}, \hat{y}, \hat{z})) + [f(u) + C] + g(\texttt{Mn}_3(\hat{x}, \hat{y}, \hat{z})) \leq f(x) + f(y) + [f(z) + C]$$

Therefore, $g(\texttt{Mj}_1(\hat{x}, \hat{y}, \hat{z})) < C$, and thus $g(\texttt{Mj}_1(\hat{x}, \hat{y}, \hat{z})) = f(b, \texttt{Mj}_1(\hat{x}, \hat{y}, \hat{z})) = f(\texttt{Mj}_1(x, y, z))$. Similarly, $g(\texttt{Mn}_3(\hat{x}, \hat{y}, \hat{z})) = f(c, \texttt{Mn}_3(\hat{x}, \hat{y}, \hat{z})) = f(\texttt{Mn}_3(x, y, z))$, and hence the inequality above is equivalent to (6).

**Case 2** $(\texttt{Mj}_2(\hat{x}, \hat{y}, \hat{z}), \hat{y}, \hat{z}) \not\prec (\hat{x}, \hat{y}, \hat{z})$. This implies, in particular, the following condition:

---

[1] If $u_j = v_j$ then obviously $\texttt{MJN}(u_j, v_j, z_j) = (u_j, v_j, z_j)$; suppose that $u_j \neq v_j$. This implies $u_j \neq x_j$ and $u_j \neq y_j$ (if $u_j = y_j$ then we would have $v_j = \texttt{Mj}_2(u_j, u_j, z_j) = u_j$). Therefore, $u_j = z_j$. We must have $v_j = \texttt{Mj}_2(z_j, y_j, z_j) = y_j$ since $v_j \neq u_j = z_j$. Thus, $\texttt{MJN}(u_j, v_j, z_j) = \texttt{MJN}(z_j, y_j, z_j) = (\alpha, y_j, \beta)$. We have $\{\{z_j, y_j, z_j\}\} = \{\{\alpha, y_j, \beta\}\}$, and so $\alpha = \beta = z_j$.

(∗) if $|\{x_j, y_j, z_j\}| = 3$ for $j \in V - \{i\}$ then $\text{Mj}_2(x_j, y_j, z_j) = x_j$.

It is easy to check that $\Delta(\text{Mj}_2(\hat{\boldsymbol{x}}, \hat{\boldsymbol{y}}, \hat{\boldsymbol{z}}), \hat{\boldsymbol{y}}, \hat{\boldsymbol{z}}) \subseteq \Delta(\hat{\boldsymbol{x}}, \hat{\boldsymbol{y}}, \hat{\boldsymbol{z}})$. Indeed, consider node $j \in V - \{i\}$ with $\text{Mj}_2(x_j, y_j, z_j) \neq y_j$; we need to show that $x_j \neq y_j$. If $|\{x_j, y_j, z_j\}| = 3$ then this follows from (∗), so it remains to consider the case when $\text{MJN}(x_j, y_j, z_j)$ is defined via (5d) (case (5a) was eliminated by proposition 24). We then have $\text{Mj}_2(x_j, y_j, z_j) = x_j \sqcup y_j$, and so $x_j \sqcup y_j \neq y_j$ clearly implies $x_j \neq y_j$.

We thus must have $\Delta(\text{Mj}_2(\hat{\boldsymbol{x}}, \hat{\boldsymbol{y}}, \hat{\boldsymbol{z}}), \hat{\boldsymbol{y}}, \hat{\boldsymbol{z}}) = \Delta(\hat{\boldsymbol{x}}, \hat{\boldsymbol{y}}, \hat{\boldsymbol{z}})$, otherwise the assumption of Case 2 would not hold. This implies the following:

(∗∗) if $x_j \neq y_j$ for $j \in V - \{i\}$ then $\text{Mj}_2(x_j, y_j, z_j) \neq y_j$.

Let us define $\boldsymbol{u} = \text{Mj}_1(\boldsymbol{x}, \boldsymbol{y}, \boldsymbol{z})$, and let $\boldsymbol{x}', \boldsymbol{u}'$ be the labelings obtained from $\boldsymbol{x}, \boldsymbol{u}$ by setting $x'_i = u'_i = z_i$, so that we have

$$\begin{array}{l} a = x_i = u_i = \text{Mj}_1(x_i, y_i, z_i) \\ b = x'_i = u'_i = \text{Mj}_2(x_i, y_i, z_i) \ (= z_i) \\ c = \text{Mn}_3(x_i, y_i, z_i) \ (= y_i) \end{array}$$

We claim that the following identities hold:

$$\begin{array}{rclcrcl} \text{Mj}_1(\boldsymbol{x}', \boldsymbol{y}, \boldsymbol{z}) & = & \boldsymbol{u}' & \quad & \boldsymbol{x} \sqcap \boldsymbol{u}' & = & \text{Mj}_1(\boldsymbol{x}, \boldsymbol{y}, \boldsymbol{z}) = \boldsymbol{u} \\ \text{Mj}_2(\boldsymbol{x}', \boldsymbol{y}, \boldsymbol{z}) & = & \text{Mj}_2(\boldsymbol{x}, \boldsymbol{y}, \boldsymbol{z}) & & \boldsymbol{x} \sqcup \boldsymbol{u}' & = & \boldsymbol{x}' \\ \text{Mn}_3(\boldsymbol{x}', \boldsymbol{y}, \boldsymbol{z}) & = & \text{Mn}_3(\boldsymbol{x}, \boldsymbol{y}, \boldsymbol{z}) & & & & \end{array}$$

Indeed, we need to show that $x_j \sqcup u_j = x_j$ for $j \in V - \{i\}$. If $\text{MJN}(x_j, y_j, z_j)$ was defined via (5b) then $\text{Mj}_2(x_j, y_j, z_j) = y_j \sqcup z_j \neq x_j$ contradicting to condition (∗). Similarly, if it was defined via (5c) then $\text{Mj}_2(x_j, y_j, z_j) = x_j \sqcup z_j = z_j \neq x_j$ again contradicting to condition (∗). (Note, in the latter case $x_j \sqcup z_j = z_j$ since by proposition 26 we cannot have $\{x_j, z_j\} \in \overline{M}$.) We showed that $\text{MJN}(x_j, y_j, z_j)$ must be determined via (5d), so $u_j = \text{Mj}_1(x_j, y_j, z_j) = x_j \sqcap y_j$ and $\text{Mj}_2(x_j, y_j, z_j) = x_j \sqcup y_j$. If $x_j = y_j$ then the claim $x_j \sqcup u_j = x_j$ is trivial. If $x_j \neq y_j$ then condition (∗∗) implies $x_j \sqcup y_j \neq y_j$, and consequently $x_j \sqcup y_j = x_j$, $u_j = x_j \sqcap y_j = y_j$ and $x_j \sqcup u_j = x_j \sqcup y_j = x_j$, as claimed.

The rest of the proof proceeds analogously to the proof for the case T1. $\square$

**Proposition 28.** *For node $i \in V$ the following situation is impossible:*

T6    $\mu(\{x_i, y_i, z_i\}) = \{x_i\}, \{y_i, z_i\} \in \overline{M}$.

*Proof.* Let us define $\boldsymbol{u} = \text{Mj}_2(\boldsymbol{x}, \boldsymbol{y}, \boldsymbol{z})$ and $\boldsymbol{v} = \text{Mj}_2(\boldsymbol{u}, \boldsymbol{x}, \boldsymbol{z})$. It can be checked that $\text{MJN}(\boldsymbol{v}, \boldsymbol{u}, \boldsymbol{z}) = (\boldsymbol{v}, \boldsymbol{u}, \boldsymbol{z})$.[2] Therefore, if we define $\boldsymbol{z}' = \boldsymbol{z}$ and $\boldsymbol{u}' = \boldsymbol{u}$ then the following identities will hold:

$$\begin{array}{rclcrclcrcl} \text{Mj}_1(\boldsymbol{x}, \boldsymbol{y}, \boldsymbol{z}') & = & \text{Mj}_1(\boldsymbol{x}, \boldsymbol{y}, \boldsymbol{z}) & \quad & \text{Mj}_1(\boldsymbol{v}, \boldsymbol{u}', \boldsymbol{z}) & = & \boldsymbol{v} & \quad & \boldsymbol{v} & = & \text{Mj}_2(\boldsymbol{u}', \boldsymbol{x}, \boldsymbol{z}) \\ \text{Mj}_2(\boldsymbol{x}, \boldsymbol{y}, \boldsymbol{z}') & = & \boldsymbol{u}' & & \text{Mj}_2(\boldsymbol{v}, \boldsymbol{u}', \boldsymbol{z}) & = & \text{Mj}_2(\boldsymbol{x}, \boldsymbol{y}, \boldsymbol{z}) = \boldsymbol{u} & & & & \\ \text{Mn}_3(\boldsymbol{x}, \boldsymbol{y}, \boldsymbol{z}') & = & \text{Mn}_3(\boldsymbol{x}, \boldsymbol{y}, \boldsymbol{z}) & & \text{Mn}_3(\boldsymbol{v}, \boldsymbol{u}', \boldsymbol{z}) & = & \boldsymbol{z}' & & & & \end{array}$$

---

[2] If $u_j = v_j$ then obviously $\text{MJN}(v_j, u_j, z_j) = (v_j, u_j, z_j)$; suppose that $u_j \neq v_j$. This implies $u_j \neq x_j$ (otherwise we would have $v_j = \text{Mj}_2(u_j, u_j, z_j) = u_j$). If $\text{MJN}(x_j, y_j, z_j)$ is determined via (5b) then $\{y_j, z_j\} \in \overline{M}$ by proposition 26 and so $u_j = z_j$ and $v_j = z_j$. It remains to consider the case when it is determined via (5d) (cases (5a) and (5c) have been eliminated).

We have $u_j = x_j \sqcup y_j = y_j$ since $u_j \neq x_j$, and so $v_j = \text{Mj}_2(y_j, x_j, z_j) = y_j \sqcup x_j = x_j$ since $v_j \neq u_j = y_j$ (clearly, $\text{Mj}_2(y_j, x_j, z_j)$ is also determined via (5d)). We thus have $\text{MJN}(v_j, u_j, z_j) = \text{MJN}(x_j, y_j, z_j) = (\alpha, u_j, z_j)$. Condition $\{\{v_j, u_j, z_j\}\} = \{\{\alpha, u_j, z_j\}\}$ implies that $\alpha = v_j$.

Let us modify $z'$ and $u'$ according to the following diagram:

$$\begin{array}{rl} a = z_i = u_i & = \mathtt{Mj}_2(x_i, y_i, z_i) \; (= v_i) \\ b = z'_i = u'_i & = \mathtt{Mj}_1(x_i, y_i, z_i) \; (= y_i) \\ c & = \mathtt{Mn}_3(x_i, y_i, z_i) \; (= x_i) \end{array}$$

It can be checked that the identities above still hold. It suffices to show that $(u', x, z) \prec (x, y, z)$, then the proof will be analogous to the proof for the Case 1 of T5.

Consider node $j \in V - \{i\}$. We will show next that $j$ satisfies the following:

(a) If $j \in \Delta(u', x, z)$ then $j \in \Delta(x, y, z)$. In other words, if $u'_j \ne x_j$ then $y_j \ne x_j$.

(b) If $j \in \Delta^M(u', x, z)$ then $j \in \Delta^M(x, y, z)$. Namely, if $(u'_j, x_j, z_j) = (a, b, b)$ or $(u'_j, x_j, z_j) = (b, a, b)$ where $\{a, b\} \in M$ then $u'_i = y_i$ and thus $(x_i, y_i, z_i) = (b, a, b)$ or $(x_i, y_i, z_i) = (a, b, b)$ respectively.

(c) $\mu(\{u'_j, x_j, z_j\}) \ne \{u'_j\}$.

This will imply the claim since $(u'_i, x_i, z_i) = (y_i, x_i, z_i) \prec (x_i, y_i, z_i)$ due to the fourth component in (7).

If $\mathtt{MJN}(x_j, y_j, z_j)$ is determined via (5b) then we must have $\{y_j, z_j\} \in \overline{M}$ by proposition 26, and so $u'_j = \mathtt{Mj}_2(x_j, y_j, z_j) = z_j$. Checking (a-c) is then straightforward.

It remains to consider the case when $\mathtt{MJN}(x_j, y_j, z_j)$ is determined via (5d) - all other cases have been eliminated. Condition (c) then clearly holds, and $u'_j = \mathtt{Mj}_2(x_j, y_j, z_j) = x_j \sqcup y_j$. If $u'_j = x_j$ then (a,b) are trivial since their preconditions do not hold. It is also straightforward to check that (a,b) hold if $u'_j = y_j \ne x_j$.

$\square$

## Appendix A

In this section we prove proposition 15.

**Part (a)** One direction is trivial: if $\{(a, b), (a', b')\} \in G_\Gamma$ then $\{(a, b), (a', b')\} \in G_{\bar\Gamma}$, and if $\{(a, b), (a', b')\}$ is soft in $G_\Gamma$ then it is also soft in $G_{\bar\Gamma}$. For the other direction we need to show the following: (i) if $\{(a, b), (a', b')\}$ is an edge in $G_{\bar\Gamma}$ then it is also an edge in $G_\Gamma$, and (ii) if $\{(a, b), (a', b')\}$ is a soft edge in $G_{\bar\Gamma}$ then it is also soft in $G_\Gamma$.

Suppose that $\{(a, b), (a', b')\} \in \Gamma^*$. Let $f \in (\bar\Gamma)^*$ be the corresponding binary function. If $\{(a, b), (a', b')\}$ is soft in $\Gamma^*$, then we choose $f$ according to the definition of the soft edge. We have

$$f(x, y) = \min_{z \in D^{m-2}} g(x, y, z) \qquad \forall x, y \in D$$

where $g : D^m \to \overline{\mathbb{R}}_+$ is a sum of cost functions from $\bar\Gamma$. Let $C$ be a sufficiently large finite constant (namely, $C > \max\{g(z) \mid z \in \mathtt{dom} g\}$), and let $g^C$ be the function obtained from $g$ as follows: we take every unary cost function $u : D \to \overline{\mathbb{R}}_+$ present in $g$ and replace it with function $u^C(z) = \min\{u(z), C\}$. Clearly, $g^C \in \Gamma^*$. Define

$$f^C(x, y) = \min_{z \in D^{m-2}} g^C(x, y, z) \qquad \forall x, y \in D$$

then $f^C \in \Gamma^*$. It is easy to see that $f$ and $f^C$ have the following relationship: (i) if $f(x, y) < \infty$ then $f^C(x, y) = f(x, y) < C$; (ii) if $f(x, y) = \infty$ then $f^C(x, y) \ge C$. We have $f(a, a') + f(b, b') > f(a, b') + f(b, a')$ and $(a, b'), (b, a') \in \mathtt{dom} f$; this implies that $f^C(a, a') + f^C(b, b') > f^C(a, b') + f^C(b, a')$, and thus $\{p,q\} \in G_\Gamma$. If edge $\{p,q\}$ is soft in $G_{\bar\Gamma}$ then at least one of the assignments $(a,a'),(b,b')$ is in $\text{dom}\, f$ (and thus in $\text{dom}\, f^C$), and so $\{p,q\}$ is soft in $G_\Gamma$.

**Part (b)** Suppose that $\bar\Gamma$ is NP-hard, i.e. there exists a finite language $\bar\Gamma' \subseteq \bar\Gamma$ which is NP-hard. Define $C = \max\{f(\boldsymbol{x}) \mid f \in \bar\Gamma', \boldsymbol{x} \in \text{dom}\, f\} + 1$. Let $\Gamma'$ be the language obtained from $\bar\Gamma'$ as follows: we take every unary cost function $u : D \to \overline{\mathbb{R}}_+$ present in $\bar\Gamma'$ and replace it with function $u^C(z) = \min\{u(z), C\}$. Clearly, $\Gamma' \subseteq \Gamma$. We prove below that $\Gamma'$ is NP-hard using a reduction from $\bar\Gamma'$.

Let $\bar{\mathcal{I}}$ be an instance from $\bar\Gamma'$ with the cost function
$$f(\boldsymbol{x}) = \sum_{t \in T_1} u_t\left(x_{i(t,1)}\right) + \sum_{t \in T_*} f_t\left(x_{i(t,1)}, \ldots, x_{i(t,m_t)}\right)$$
where $T_1$ is the index set of unary cost functions and $T_*$ is index set of cost functions of higher arities. Note, $u_t \in \bar\Gamma'$ for $t \in T_1$ and $f_t \in \bar\Gamma'$ for $t \in T_*$. Now define instance $\mathcal{I}$ with the cost function
$$f^C(\boldsymbol{x}) = \sum_{t \in T_1} N \cdot u_t^C\left(x_{i(t,1)}\right) + \sum_{t \in T_*} f_t\left(x_{i(t,1)}, \ldots, x_{i(t,m_t)}\right)$$
where $N = |T_1| + |T_*|$. It can be viewed as an instance from $\Gamma'$, if we simulate multiplication of $N$ and $u_t^C$ by repeating the latter term $N$ times; the size of the expression grows only polynomially. It is easy to see that $f$ and $f^C$ have the following relationship: (i) if $f(\boldsymbol{x}) < \infty$ then $f^C(\boldsymbol{x}) = f(\boldsymbol{x}) < N \cdot C$; (ii) if $f(\boldsymbol{x}) = \infty$ then $f^C(\boldsymbol{x}) \geq N \cdot C$. Thus, solving $\mathcal{I}$ will also solve $\bar{\mathcal{I}}$.

## Appendix B

In this section we prove the following fact: if $MinHom(\Gamma)$ is NP-hard then so is $\Gamma$.

Let $MinHom(\Gamma)' \subseteq MinHom(\Gamma)$ be a finite language with costs in $\mathbb{Z}_+ \cup \{\infty\}$ which is NP-hard. Denote $MinHom(\Gamma)'_1$ and $MinHom(\Gamma)'_*$ to be the subsets of $MinHom(\Gamma)'$ of arity $m = 1$ and $m \geq 2$ respectively. The definition of $MinHom(\Gamma)$ implies that for every $f \in MinHom(\Gamma)'_*$ there exists function $f^\circ \in \Gamma$ such that $f(\boldsymbol{x}) = 0$ if $f^\circ(\boldsymbol{x}) < \infty$, and $f(\boldsymbol{x}) = \infty$ if $f^\circ(\boldsymbol{x}) = \infty$. Denote $C = \max\{f^\circ(\boldsymbol{x}) \mid f \in MinHom(\Gamma)'_*, \boldsymbol{x} \in \text{dom}\, f^\circ\} + 1$. Construct language $\Gamma'$ as follows:
$$\Gamma' = \{u^C \mid u \in MinHom(\Gamma)'_1\} \cup \{f^\circ \mid f \in MinHom(\Gamma)'_*\}$$
where function $u^C$ is defined by $u^C(z) = C \cdot u(z)$. Clearly, $\Gamma' \subseteq \Gamma$. We prove below that $\Gamma'$ is NP-hard using a reduction from $\bar\Gamma'$.

Let $\hat{\mathcal{I}}$ be an instance from $MinHom(\Gamma)'$ with the cost function
$$f(\boldsymbol{x}) = \sum_{t \in T_1} u_t\left(x_{i(t,1)}\right) + \sum_{t \in T_*} f_t\left(x_{i(t,1)}, \ldots, x_{i(t,m_t)}\right)$$
where $T_1$ is the index set of unary cost functions and $T_*$ is index set of cost functions of higher arities. Note, $u_t \in MinHom(\Gamma)'_1$ for $t \in T_1$ and $f_t \in MinHom(\Gamma)'_*$ for $t \in T_*$. Now define instance $\mathcal{I}$ with the cost function
$$f^C(\boldsymbol{x}) = \sum_{t \in T_1} N \cdot u_t^C\left(x_{i(t,1)}\right) + \sum_{t \in T_*} f_t^\circ\left(x_{i(t,1)}, \ldots, x_{i(t,m_t)}\right)$$
where $N = |T_*|$. It can be viewed as an instance from $\Gamma'$, if we simulate multiplication of $N$ and $u_t^C$ by repeating the latter term $N$ times; the size of the expression grows only polynomially. For any $\boldsymbol{x} \in \text{dom}\, f$ we have
$$f^C(\boldsymbol{x}) \geq \sum_{t \in T_1} N \cdot u_t^C\left(x_{i(t,1)}\right) = NC \cdot f(\boldsymbol{x})$$
$$f^C(\boldsymbol{x}) < \sum_{t \in T_1} N \cdot u_t^C\left(x_{i(t,1)}\right) + \sum_{t \in T_*} C = NC \cdot (f(\boldsymbol{x}) + 1)$$

Furthermore, $f(\boldsymbol{x}) = \infty$ iff $f^C(\boldsymbol{x}) = \infty$. Function $f$ have values in $\mathbb{Z}_+ \cup \{\infty\}$, therefore solving $\mathcal{I}$ will also solve $\hat{\mathcal{I}}$.